\documentclass[aps,graphicx,twocolumn]{revtex4}

\usepackage{graphicx}
\usepackage{epstopdf}
\usepackage{verbatim}
\usepackage{amsmath}
\usepackage{amssymb}
\usepackage[colorlinks,linkcolor=blue,citecolor=blue,urlcolor=blue]{hyperref}
\usepackage[caption=false,font=scriptsize,labelfont=sf,textfont=sf]{subfig}
\usepackage{multirow}

\begin{document}
\title{Robust logical Bell nonlocality based on quantum error correction codes}

\author{Qi Zhang$^{1,2}$, Jia-Wei Ying$^{3}$, Cheng Liu$^{1}$, Lan Zhou$^{1}$\footnote{Email address: zhoul727@hznu.edu.cn}, Yu-Bo Sheng$^{1,3}$\footnote{Email address: shengyb@njupt.edu.cn}}
\address{$^1$School of Physics, Hangzhou Normal University, Hangzhou, Zhejiang 311121, China\\
$^2$College of Science, Nanjing University of Posts and Telecommunications, Nanjing, Jiangsu 210023, China\\
$^3$College of Electronic and Optical Engineering and College of Flexible Electronics (Future Technology), Nanjing University of Posts and Telecommunications, Nanjing, Jiangsu 210023, China\\
}
\date{\today}

\begin{abstract}
Quantum nonlocality based on the violation of Bell-like inequalities constitutes a fundamental feature of quantum physics and drives the development of device-independent (DI) quantum information technologies. Existing studies of Bell nonlocality have mainly focused on physical qubit systems, where the observed nonlocal correlations are directly encoded in physical degrees of freedom. The decoherence sensitivity of Bell nonlocality largely limits the performance and security of its DI applications. Here, we investigate the robust logical Bell nonlocality based on quantum error correction codes. We construct the general logical Bell inequality in the stabilizer coding subspace and prove its violation indicates the global nonlocal feature of the logical system. Then, we indicate that the logical Bell nonlocality is robust against decoherence. Comparing with the physical qubit system, the fidelity thresholds for the logical Bell inequality violation based on the [[3,1,1]] and [[7,1,1]] repetition codes under the bit-flip error model can be reduced from 82.8\% to 73.10\% and 66.35\%, increasing DI QKD's bit-flip noise threshold from 10.64\% to 14.42\% and 23.36\%, respectively. Such stabilizer-based framework can be also used to characterize the multipartite logical Bell nonlocality in principle. Finally, a logical Bell test implementation circuit based on the [[3,1,1]] repetition code is presented. This work provides a feasible avenue for unlocking robust Bell nonlocality in scalable logical quantum systems and facilitates its applications in future scalable quantum network.
\end{abstract}
\maketitle

Quantum nonlocality is the most representative nonclassical features of quantum mechanics \cite{Brunner,NN}. Bell's theorem constitutes a theoretical criterion that the violation of a Bell-like inequality is sufficient to demonstrate the quantum nonlocality \cite{Bell,CHSH}. The exploration of Bell-like inequalities violation not only contributes to the foundation of quantum mechanics but also drives the development of quantum information technologies, such as the device-independent (DI) quantum communication protocols \cite{DIQKD2,DIQSDC1,DIQSS1}, DI random number generation \cite{RNC} and self-testing \cite{selftest1,selftest2}. For example, DI quantum communication protocols have constructed quantitative relations between the security bounds and the Bell nonlocality \cite{DIQKD2,DIQSDC1,DIQSS1}. The experimental demonstrations of DI quantum key distribution (QKD) mark a key step for the Bell nonlocality in practical quantum communication utilization \cite{DIQKD e1,DIQKD e2,DIQKD e3,DIQKD e4}.

The Bell-like inequalities violations have been successfully verified on various physical platforms, including the photons \cite{Belltest1,Belltest2,Belltest3,Belltest3a}, trapped ions \cite{Belltest4}, atoms \cite{Belltest6}, and superconducting platforms \cite{Belltest5,Belltest5a}. Furthermore, loophole-free Bell tests have also been realized in photon \cite{Belltest8,Belltest9}, electron spin \cite{Belltest7} and superconducting systems \cite{Belltest5a}. Existing researches on Bell nonlocality mainly focus on physical qubit systems. The physical Bell nonlocality is sensitive to decoherence, which poses challenges for its practical applications. For instance, the decoherence raises the global detection efficiency threshold for the DI quantum communication protocols, thus largely increasing their implementation difficulty and limiting the secure communication distances \cite{DIQKD2,DIQSDC1,DIQSS1}. For depolarizing noise channel, the noise threshold of DI QKD is as low as 7.1\% \cite{DIQKD2}.

The quantum error correction (QEC) is a promising method for protecting logical qubit system against the decoherence \cite{QEC4,QEC5,QEC1,QEC6}. Compared with a physical qubit that supports only two-dimensional Hilbert space, a logical qubit is in a two-dimensional code subspace of a high-dimensional quantum system, where the redundant degrees of freedom allow the detection and correction of possible errors \cite{QEC1,QEC6,GKPlogical}. Logical qubit systems have become critical resources for future quantum networks and long-distance quantum communication \cite{repeater1,repeater2,repeater3,repeater5}. This raises two interesting questions. Can Bell nonlocality be defined and observed at the logical level? If so, does the logical layer exhibit the stronger robustness Bell nonlocality against decoherence than the physical layer?

Existing Bell inequalities explicitly rely on local measurements performed on individual physical qubits. As the physical qubit does not have the error subspace, the existing Bell inequalities for physical qubit systems cannot be directly applied for the logical Bell nonlocality characterization. In this work, we construct a general logical Bell inequality which is universally defined for the logical systems based on QEC codes. The logical Bell inequality violation indicates the global nonlocal feature of the logical system in the stabilizer coding subspace. We prove that the logical encoding can enhance the logical Bell nonlocality's noise robustness. Our work provides a feasible avenue for unlocking robust Bell nonlocality in scalable logical quantum systems, and facilitates its applications in future scalable quantum network.

Before introducing the logical Bell inequality, we first recall the physical content of general Bell inequalities. Bell inequalities constrain whether the observed correlation can be explained by a local hidden-variable (LHV) model. This constraint does not depend on the specific physical implementation of the system, but solely on the following fundamental assumptions \cite{Bell}:

1. The system consists of two spatially or operationally distinguishable subsystems A and B. The whole Hilbert space $\mathcal{H}$ admits a tensor-product decomposition $\mathcal{H}=\mathcal{H}_A\otimes\mathcal{H}_B$;

2. The subsystems A and B can choose among several local observables $A_x$ and $B_y$ ($x$, $y \in\{0,1\}$), with measurement outcomes of $\pm1$, respectively;

3. All measurement statistics are given by the expected values of these observables with respect to the quantum state.

Under these assumptions, the violation of the Bell (Clauser-Horne-Shimony-Holt, CHSH) inequality certifies the presence of quantum nonlocality between the subsystems A and B.

Logical qubits are embedded via isometric encoding into a code subspace within a higher-dimensional Hilbert space. The corresponding logical observables are characterized by a family of physical operators equivalent in the code subspace.

\textbf{Proposition.} Consider an arbitrary $[[n,1]]$ stabilizer code with the code Hilbert space $\mathcal{H}_{\mathrm{code}}=\mathrm{span}\{|0_L\rangle,|1_L\rangle\} \subset \mathcal{H}_{\mathrm{phy}}=(\mathbb{C}^2)^{\otimes n}$, where $|0_L\rangle$ and $|1_L\rangle$ are logical computational basis. The logical measurement operators from A and B are defined as
\begin{eqnarray}\label{measurement operator}
\bar A_0&=&\bar{Z}=|0_L\rangle\langle0_L|-|1_L\rangle\langle1_L|, \quad \bar B_0=\frac{\bar{Z}+\bar{X}}{\sqrt{2}},\nonumber\\
\bar A_1&=&\bar{X}=|0_L\rangle\langle1_L|+|1_L\rangle\langle0_L|, \quad \bar B_1=\frac{\bar{Z}-\bar{X}}{\sqrt{2}},
\end{eqnarray}
satisfying $\{\bar{Z},\bar{X}\}=0$ and $\bar{X}^2=\bar{Z}^2=\mathbb{I}_L=|0_L\rangle\langle0_L|+|1_L\rangle\langle1_L|$. The corresponding logical CHSH operator is defined as
\begin{eqnarray}\label{CHSH operator}
\mathcal{B}_L=\bar A_0\otimes \bar B_0+\bar A_0\otimes \bar B_1+\bar A_1\otimes \bar B_0-\bar A_1\otimes \bar B_1.
\end{eqnarray}
For any LHV model, the logical CHSH inequality $\langle\mathcal{B}_L\rangle\le2$ must be satisfied. If the subsystems exhibit nonlocal correlation, the maximal achievable expected value of $\mathcal{B}_L$ can reach $2\sqrt{2}$, which saturates the Tsirelson bound \cite{Tsirelson}.

\textbf{Proof.} Any $[[n,1]]$ stabilizer code can be mathematically described as an isometric mapping
\begin{eqnarray}\label{isometric mapping}
V=|0_L\rangle\langle 0|+|1_L\rangle\langle 1|:\mathbb{C}^2\rightarrow(\mathbb{C}^2)^{\otimes n},\quad V^\dagger V=\mathbb{I},
\end{eqnarray}
which maps an individual physical qubit into an $n$ qubit code subspace.

Here, we clarify the reasons why the general physical CHSH inequalities are not applicable at the logical level with above stabilizer code structure. First, for a bipartite system shared by Alice and Bob, the encoding operations act locally as $V_A$ and $V_B$, respectively. $V_A$ and $V_B$ can map an arbitrary two-qubit physical state $|\phi\rangle$ to a logical bipartite state
\begin{eqnarray}\label{logical state}
|\phi_L\rangle=(V_A\otimes V_B)|\phi\rangle.
\end{eqnarray}
This construction preserves locality at the encoding mapping level.

Within the code subspace $\mathcal{H}_{\mathrm{code}}$, the logical operator $\bar O$ corresponding to the physical operator $O$ is defined as $\bar{O}=VOV^\dagger$, with $O\in\{A_x, B_y\}$ ($x$, $y \in\{0,1\}$). It can be found that logical operators are not simply associated with individual physical subsystems or their tensor-product structure. This is one reason why physical Bell inequalities cannot be directly applied to characterize the logical Bell nonlocality.

Next, although logical Bell states are physically realized as multiparticle entangled states, the joint probability distribution of the logical observables and physical observables are quite different. Under the locality assumptions, the joint probability distribution of a physical multiparticle system admits an LHV decomposition as \cite{multiparticle}
\begin{eqnarray}\label{LHV1}
&&p(\{a_i\},\{b_j\}|x,y)\nonumber\\
&&=\int d\lambda p(\lambda) \prod_i p(a_i|x,\lambda)\prod_j p(b_j|y,\lambda),
\end{eqnarray}
where $\{a_i\}$ and $\{b_j\}$ denote the measurement outcome sets of Alice and Bob in their physical subsystems, respectively, and each physical subsystem is associated with an independent local response function. By contrast, in a logical Bell state scenario, measurement outcomes of logical observables are defined within an encoded subspace. The logical outcomes are obtained as a functions $f$ of multiple physical measurement outcomes as
\begin{eqnarray}
\bar{a} = f(a_1,\dots,a_n), \qquad \bar{b} = f(b_1,\dots,b_n).
\end{eqnarray}
In this way, the joint probability distribution of logical outcomes has the form of
\begin{eqnarray}\label{LHV2}
p(\bar{a},\bar{b}|x,y)=\int d\lambda p(\lambda) p(\bar{a}|x,\lambda)p(\bar{b}|y,\lambda),
\end{eqnarray}
which is fundamentally different from Eq.~\eqref{LHV1}. This difference also makes the physical Bell inequalities cannot be directly applied to characterize the logical nonlocality.

Then, we construct the logical CHSH operator. For the logical bipartite state $|\phi_L\rangle$, the expected value of logical measurement operators $\bar A_x \otimes \bar B_y$ can be calculated as
\begin{eqnarray}\label{Expected value}
&&\langle \phi_L|\bar A_x \otimes \bar B_y|\phi_L \rangle \nonumber\\
&&= \langle \phi|(V_A^\dagger \otimes V_B^\dagger)(\bar{A}_x \otimes \bar{B}_y)(V_A \otimes V_B)|\phi \rangle \nonumber\\
&&= \langle \phi| A_x \otimes B_y |\phi \rangle,
\end{eqnarray}
where $A_x = V^\dagger_A\bar{A}_xV_A$ and $B_y = V^\dagger_B\bar{B}_yV_B$ ($x$, $y \in\{0,1\}$). It can be found that all statistics based on the expected values and correlation functions remain invariant before and after the logical encoding. Therefore, the logical isometric mapping essentially embeds the Hilbert space $\mathcal{H}_{\mathrm{code}}$ of a logical qubit into a higher-dimensional physical Hilbert space $\mathcal{H}_{\mathrm{phy}}$, while strictly preserving the entanglement structure of the quantum state.

It is noticed that within the code subspace $\mathcal{H}_{\mathrm{code}}$, the logical measurement operators satisfy $\bar{A}_x^2=\bar{B}_y^2=\mathbb{I}_L$ ($x$, $y \in\{0,1\}$). Their spectra consist solely of the eigenvalues $\pm1$, and the corresponding logical measurement outcomes are $\bar{a}_x,\bar{b}_y\in\{+1,-1\}$. Therefore, based on the locality assumption, the expected value of $\bar{a}_x$ and $\bar{b}_y$ obtained under measurement choices $x$ and $y$ is defined as
\begin{eqnarray}\label{Expected value2}
E(\bar{a}_x \bar{b}_y)&=&\langle \bar{A}_x \otimes \bar{B}_y \rangle_\lambda =\sum_{\bar{a},\bar{b}} \bar{a}\bar{b}p(\bar{a},\bar{b}|x,y,\lambda) \nonumber\\
&=&\int d\lambda p(\lambda)\langle \bar{A}_x \rangle_\lambda \langle \bar{B}_y \rangle_\lambda,
\end{eqnarray}
where $\langle \bar{A}_x \rangle_\lambda=\sum_{\bar{a}} \bar{a}p(\bar{a}|x,\lambda)$ and
$\langle \bar{B}_y \rangle_\lambda=\sum_{\bar{b}} \bar{b}p(\bar{b}|y,\lambda)$
are in the range of $[-1,1]$. Consequently, the expected value of the logical CHSH operator satisfies
\begin{eqnarray}\label{CHSH2}
\langle \mathcal{B}_L \rangle&=&\int d\lambda p(\lambda)\big(\langle \bar{A}_0 \rangle_\lambda \langle \bar{B}_0 \rangle_\lambda+\langle \bar{A}_0 \rangle_\lambda \langle \bar{B}_1 \rangle_\lambda \nonumber\\
&+&\langle \bar{A}_1 \rangle_\lambda \langle \bar{B}_0 \rangle_\lambda-\langle \bar{A}_1 \rangle_\lambda \langle \bar{B}_1 \rangle_\lambda\big) \nonumber\\
&\le& \int d\lambda p(\lambda)\left(\left|\langle \bar{B}_0 \rangle_\lambda + \langle \bar{B}_1\rangle_\lambda \right|+\left| \langle \bar{B}_0 \rangle_\lambda - \langle \bar{B}_1\rangle_\lambda \right|\right) \nonumber\\
&\le& 2 \int d\lambda p(\lambda) \left| \langle \bar{B}_0 \rangle_\lambda \right|\le 2.
\end{eqnarray}
The above LHV model relies only on the binary nature of the measurement outcomes and the locality assumption, independent of the logical encoding structure. Therefore, it holds equally for any logical operator.

Since the logical CHSH operator $\mathcal{B}_L$ is Hermitian and the binary logical observables satisfy $\|\bar{A}_x\|\le 1$ and $\|\bar{B}_y\|\le 1$, we have
\begin{eqnarray}\label{norm}
\|\mathcal{B}_L^2\|&=&\|\left(\bar{A}_0\otimes(\bar{B}_0+\bar{B}_1)+\bar{A}_1\otimes(\bar{B}_0-\bar{B}_1)\right)^2\|\nonumber\\
&=&\|\bar{A}_0^2\otimes(\bar{B}_0+\bar{B}_1)^2+\bar{A}_1^2\otimes(\bar{B}_0-\bar{B}_1)^2 \nonumber\\
&-&[\bar{A}_0,\bar{A}_1]\otimes[\bar{B}_0,\bar{B}_1]\| \nonumber\\
&\le& 4+\|[\bar{A}_0,\bar{A}_1]\|\|[\bar{B}_0,\bar{B}_1]\|\le 8.
\end{eqnarray}
The operator norm of the logical CHSH operator satisfies $\|\mathcal{B}_L\|=\sqrt{\|\mathcal{B}_L^2\|}\le 2\sqrt{2}$. It reaches the Tsirelson bound \cite{Tsirelson} and is in contradiction with the LHV bound in Eq.~\eqref{CHSH2}.

In fact, as shown by Eq.~\eqref{Expected value}, all correlation functions constituting the CHSH operator remain identical before and after encoding. By linearly combining the above equations, we obtain
\begin{eqnarray}\label{CHSH3}
\langle \mathcal{B}_L \rangle&=&\mathrm{Tr}\left(\mathcal{B}_L|\Phi_L^+ \rangle\langle\Phi_L^+|\right) = \langle \Phi_L^+| \mathcal{B}_L |\Phi_L^+ \rangle \nonumber\\
&=& \langle \Phi^+| \mathcal{B}|\Phi^+\rangle = 2\sqrt{2},
\end{eqnarray}
where $|\Phi_L^+\rangle =\frac{1}{\sqrt{2}}(|0_L\rangle|0_L\rangle + |1_L\rangle|1_L\rangle)$ denotes the maximal logical Bell state. Thus, the logical CHSH polynomial can saturate the maximal value allowed by quantum mechanics.

The above proof is not limited to any specific type of QEC code. Its conclusion equally applies to all stabilizer codes scenarios, such as the repetition codes \cite{repetition}, phase codes \cite{repetition,shor code}, Shor codes \cite{shor code} and Calderbank-Shor-Steane (CSS) codes \cite{CSS1,CSS2}. Since the stabilizer codes encoding constitutes a local isometric mapping that preserves the tensor-product structure of the Hilbert space, the operator algebra, the measurement statistics and the logical observables defined on the code subspace $\mathcal{H}_{\mathrm{code}}$ are mathematically isomorphic to physical observables. Therefore, the corresponding logical CHSH polynomials possess the same local bound and quantum bound as their physical counterparts, respectively. In the case of $n=1$, the physical CHSH inequality can be contained within the general framework of the logical CHSH inequality.

We now discuss the significance of the logical CHSH inequality in the presence of noise. As the fundamental QEC code, the repetition code can efficiently correct bit errors. It serves as a basic building block of the Shor code \cite{shor code} and a simplified one-dimensional variant of the surface codes \cite{surface code}. Here, we choose the simplest $[[3,1,1]]$ repetition code for investigating the noise robustness of the logical Bell nonlocality.

The $[[3,1,1]]$ repetition code encodes three physical qubits into a logical qubit. The stabilizer group $\mathcal{S}$ is generated by
\begin{eqnarray}\label{stabilizer}
S_1 = Z_1 \otimes Z_2 \otimes \mathbb{I}_3,\quad S_2 = \mathbb{I}_1 \otimes Z_2 \otimes Z_3.
\end{eqnarray}
The code subspace $\mathcal{H}_{\mathrm{code}}=\{|\psi\rangle:S_i|\psi\rangle=|\psi\rangle, \forall S_i\in\mathcal{S}\}$ ($i=1,2$). Here, $|0_L\rangle=|000\rangle$ and $|1_L\rangle=|111\rangle$ are the common eigenstate of $S_1$ and $S_2$ with eigenvalue $+1$. We observe that $|0_L\rangle$ and $|1_L\rangle$ are also eigenstates of $\bar{Z}$ with eigenvalues $\pm1$, respectively. The states $|+_L\rangle=(|0_L\rangle+|1_L\rangle)/\sqrt{2}$ and $|-_L\rangle=(|0_L\rangle-|1_L\rangle)/\sqrt{2}$ are eigenstates of $\bar{X}$ with eigenvalues $\pm 1$, respectively.

Since phase errors induced by dephasing noise and bit errors caused by energy relaxation originate from distinct physical noise mechanisms, they can be regarded as independent of each other in general. We consider a refined noise model, in which each physical qubit has a bit fidelity $F_b$ and a phase fidelity $F_p$. Since the $[[3,1,1]]$ repetition code can correct at most a single bit error but cannot correct the phase error, the logical bit fidelity and logical phase fidelity are $F_{bL}=F_b^3+3F_b^2(1-F_b)$ and $F_{pL}=F_p^3+3F_p(1-F_p)^2$, respectively. In this way, the target logical Bell state $|\Phi_L^{+}\rangle$ will degrade to the mixed state
\begin{eqnarray}\label{rhonoise}
\rho_L&=&F_b^BF_p^B|\Phi_L^+\rangle\langle \Phi_L^+| +(1-F_b^B)(1-F_p^B)|\Psi_L^-\rangle\langle \Psi_L^-| \nonumber\\
&+&(1-F_b^B)F_p^B|\Psi_L^+\rangle\langle \Psi_L^+| +F_b^B(1-F_p^B)|\Phi_L^-\rangle\langle \Phi_L^-|,\nonumber\\
\end{eqnarray}
where $|\Phi_L^{-}\rangle=\frac{1}{\sqrt{2}}(|0_L\rangle|0_L\rangle-|1_L\rangle|1_L\rangle)$, and $|\Psi_L^{\pm}\rangle=\frac{1}{\sqrt{2}}(|0_L\rangle|1_L\rangle \pm |1_L\rangle|0_L\rangle)$. In the logical mixed state, the logical bit fidelity $F_b^B$ and logical phase fidelity $F_p^B$ can be calculated as
\begin{eqnarray}\label{fidelity L}
F_b^B &=& (F_{bL})^2+(1-F_{bL})^{2}, \nonumber\\
F_p^B &=& (F_{pL})^2+(1-F_{pL})^{2}.
\end{eqnarray}

Therefore, the logical CHSH polynomial ($S_{CHSH}$) in this noise model is
\begin{eqnarray}\label{CHSH noise}
S_{CHSH}=\mathrm{Tr}\left(\mathcal{B}_L\rho_L\right)= 2\sqrt{2}\left(F_b^B+F_p^B-1\right).
\end{eqnarray}

\begin{figure}
\includegraphics[width=0.46\textwidth]{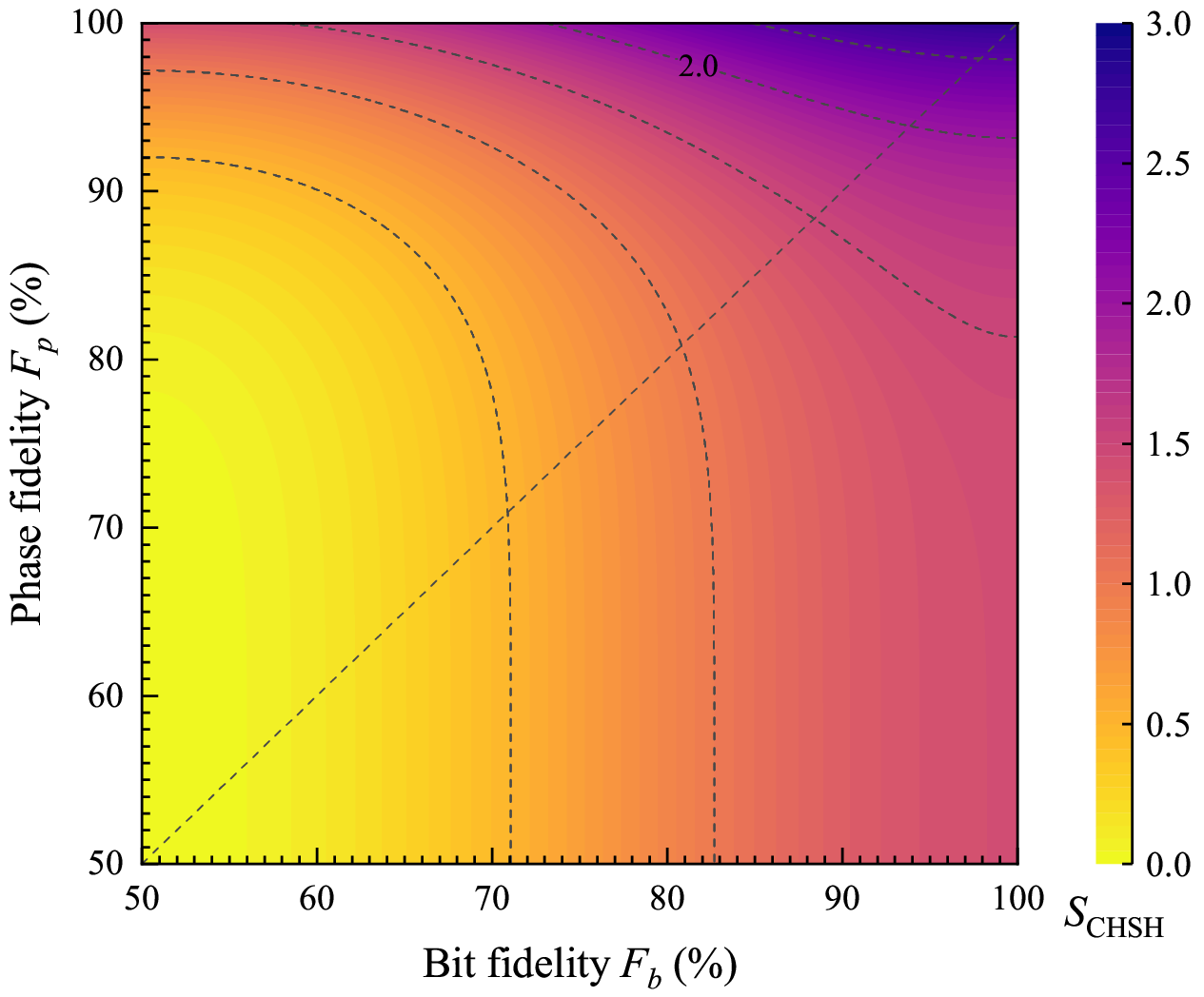}
\caption{The logical CHSH polynomial $S_{CHSH}$ based on $[[3,1,1]]$ repetition code alter with the physical bit fidelity $F_b$ and physical phase fidelity $F_p$.}
\label{fig1}
\end{figure}

Fig.~\ref{fig1} provides $S_{CHSH}$ altered with the physical bit fidelity $F_b$ and physical phase fidelity $F_p$ in the logical Bell test based on $[[3,1,1]]$ repetition code. For achieving $S_{CHSH}>2$, the thresholds of $F_b$ and $F_p$ are 73.10\% and 93.20\%, respectively, while those in the physical encoding system are both 82.20\%.
This result indicates that the logical Bell nonlocality exhibits a stronger robustness against bit error but weaker robustness against phase error.
This behavior originates from the intrinsic asymmetry of $[[3,1,1]]$ code's error correction capability. In particular, the phase error would accumulate due to the logical encoding. In the Supplementary Material, we investigate the corresponding fidelity thresholds for retaining the logical Bell nonlocality in the common stabilizer codes scenarios, including the $[[3,1,1]]$ phase code, $[[7,1,1]]$ repetition (phase) code, $[[7,1,3]]$ CSS code and $[[9,1,3]]$ Shor code. For the [[7,1,1]] repetition (phase) code, the thresholds of $F_b$ and $F_p$ are 66.35\% and 96.95\% (96.95\% and 66.35\%), respectively. For the $[[7,1,3]]$ CSS code, $F_b$ and $F_p$ thresholds are both 87.65\%, and for the $[[9,1,3]]$ Shor code, those thresholds are 84.05\% and 88.70\%, respectively.

\begin{figure}
\includegraphics[width=0.44\textwidth]{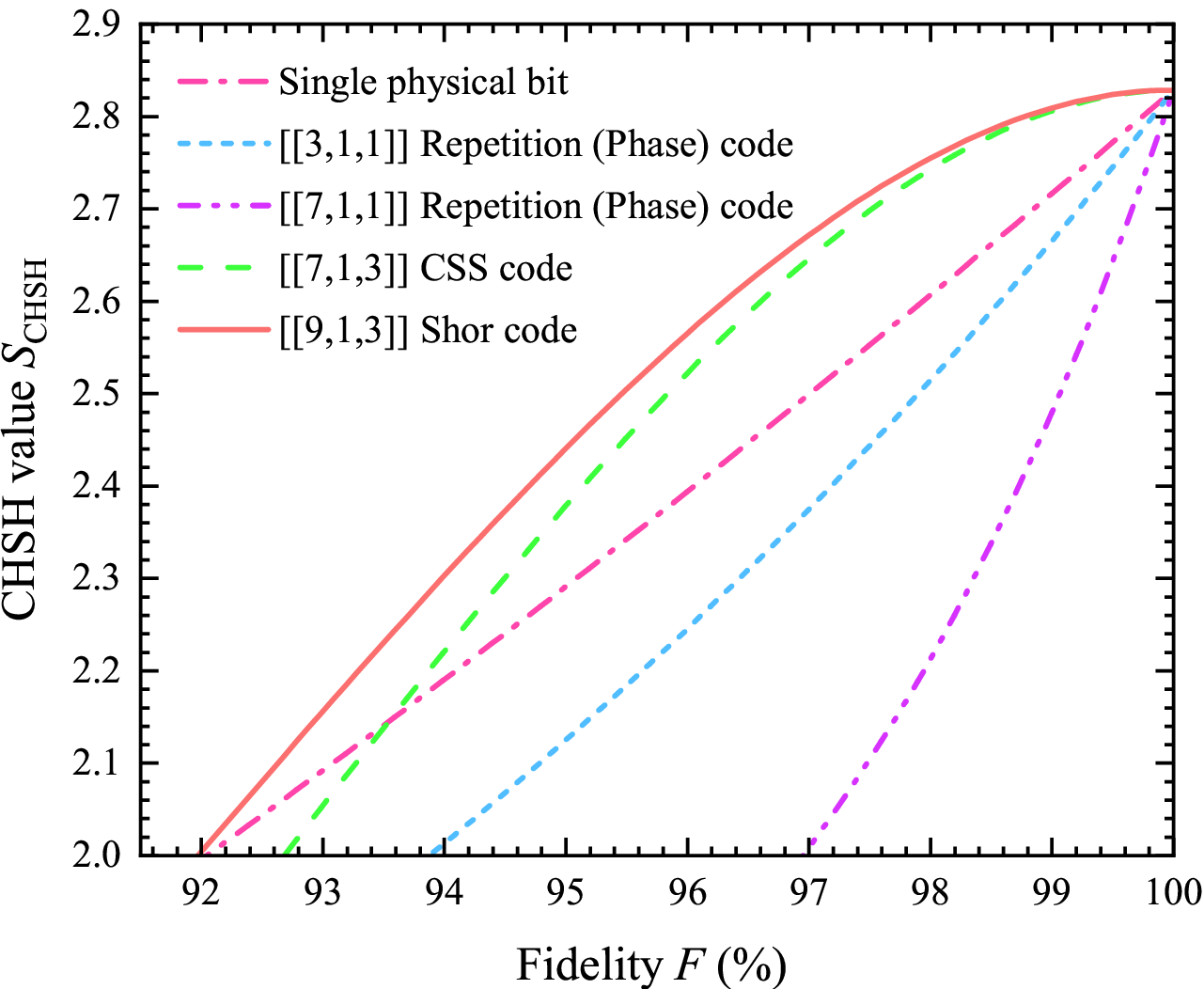}
\caption{The logical CHSH polynomial $S_{CHSH}$ based on a series of stabilizer codes altered with the fidelity $F$ under symmetric noise.}
\label{fig2}
\end{figure}

In Fig.~\ref{fig2}, we consider the symmetric noise scenario with $F_b=F_p=F$, and compare the values of $S_{CHSH}$ corresponding to a series of common QEC codes with that of the physical encoding system. Since the $[[7,1,3]]$ CSS code and $[[9,1,3]]$ Shor code can correct arbitrary single-qubit error, these cases (red solid line and green dashed line) exhibit stronger CHSH inequality violation than the physical encoding case (pink dash dot line). However, in the low-fidelity scenario ($F<93.5\%$), the logical CHSH violation based on the $[[7,1,3]]$ CSS code is slightly weaker than that of physical encoding case. This is because the multi-qubit error becomes common in the low-fidelity scenario. The stabilizer syndrome no longer reliably identifies the underlying physical errors, and the subsequent recovery operations may introduce additional logical errors, thereby suppressing the nonlocal correlations. The $F$ thresholds for retaining the nonlocality corresponding to $[[9,1,3]]$ Shor code, $[[7,1,3]]$ CSS code, and the physical encoding are about 91.98\%, 92.68\% and 92.08\%, respectively. In contrast, repetition codes and phase codes can only correct a single type (bit or phase) of error, which generally leads to weaker robustness than that of physical encoding case.

From the numerical simulations, the protection capacity of the stabilizer codes on the logical Bell nonlocality depends sensitively on the error model and the encoding structure. In the asymmetric error model, the logical nonlocality based on the asymmetric stabilizer codes, such as the repetition code or phase code has stronger robustness, while in the symmetric error model, the logical nonlocality based on symmetric stabilizer codes, such as CSS codes or Shor codes has superior robustness. For example, comparing with the physical encoding case, the fidelity thresholds for retaining the logical nonlocality based on $[[3,1,1]]$ and $[[7,1,1]]$ repetition codes under the common bit-flip error model can be reduced from 82.8\% to 73.10\% and 66.35\%, respectively, which can increase the noise threshold of DI QKD from 10.64\% \cite{DIQKD2} to 14.42\% and 23.36\%, respectively.

The stabilizer-based framework considered here is not restricted to the bipartite case and can be adopted for the multipartite logical Bell nonlocality in principle. In addition, the scale of the stabilizer code also influences the logical Bell nonlocality. For example, although the repetition (phase) code encoded with more physical bits can correct more errors, encoding a large number of physical bits would accumulate the physical bit (phase) errors. Meanwhile, the imperfect error identifying and error correction operations \cite{QEC4,QEC1,QEC5} would also accumulate with the growth of the scale. Both factors will suppresses the logical nonlocal correlations. In this way, the optimal scale of logical qubits which can maximize the protection for the corresponding logical nonlocality is an interesting topic.

Here, we take the simplest $[[3,1,1]]$ repetition code as an example to discuss the implementation circuit of the logical Bell test process.

\begin{figure*}
\includegraphics[width=0.9\textwidth]{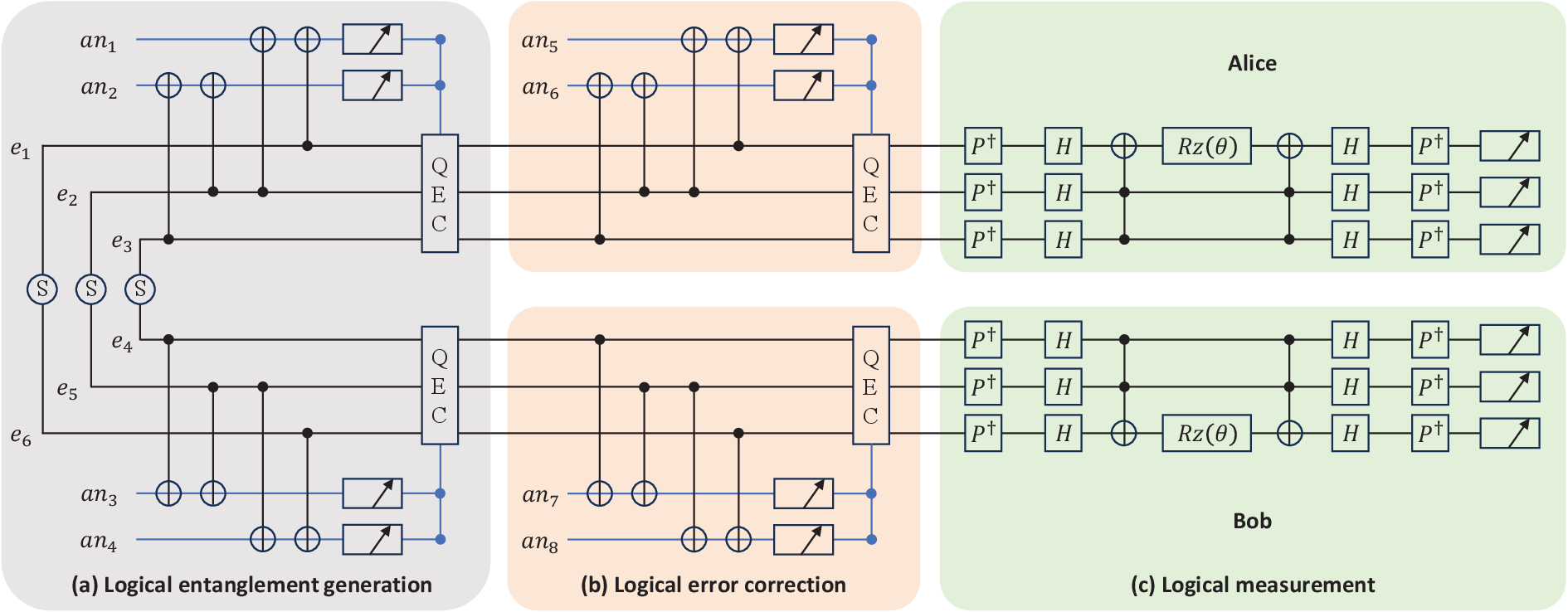}
\caption{The implementation circuit of the logical Bell test for $|\Phi_L^+\rangle =\frac{1}{\sqrt{2}}(|0_L\rangle|0_L\rangle + |1_L\rangle|1_L\rangle)$ based on the $[[3,1,1]]$ repetition code, where $|0_{L}\rangle=|000\rangle$ and $|1_{L}\rangle=|111\rangle$. The logical Bell test includes (a) logical entanglement generation, (b) logical error correction and (c) logical measurement. The circuit requires three pairs of physical Bell state $|\Phi^+\rangle=\frac{1}{\sqrt{2}}(|0\rangle|0\rangle + |1\rangle|1\rangle)$ $e_1-e_6$ and four anciliary qubits in $|0\rangle$ to construct $|\Phi_L^+\rangle$, and four auxiliary physical qubits in $|0\rangle$ to assist the error correction. The circuit is constructed by the controlled-not (CNOT) gates, the phase (P) gates, Hadamard (H) gates, and rotation gates ($R_z(\theta)$). The logical measurements are transformed to the product of physical measurements.}
\label{fig3}
\end{figure*}

The circuit of the logical Bell test for $|\Phi_L^+\rangle$ based on the $[[3,1,1]]$ repetition code is shown in Fig.~\ref{fig3}. The logical Bell state $|\Phi_L^+\rangle$ can be generated with six physical qubits $e_1$-$e_6$ in three pairs of physical Bell states $|\Phi^+\rangle=\frac{1}{\sqrt{2}}(|0\rangle|0\rangle + |1\rangle|1\rangle)$ (Fig.~\ref{fig3}(a)) \cite{Jiang}. The logical entanglement source performs the stabilizer measurements according to the selected stabilizer code by introducing ancillary qubits $an_1$-$an_4$ and extracting the corresponding error syndromes through the controlled-not (CNOT) operations. The stabilizer measurements project the whole initial physical Bell state $|\Phi^+\rangle^{\otimes 3}$ into a logical entangled state associated with the measured syndrome outcomes. Subsequently, the logical entanglement source applies the corresponding QEC operations to recover the logical entangled state into $|\Phi_L^+\rangle$.

In the error correction stage (Fig.~\ref{fig3}(b)), at each party's location, two auxiliary physical qubits in $|0\rangle$ interact with the three encoding physical qubits through the CNOT gates and then are measured in the $Z$ basis. The parties can identify the error based on the measurement results and actively recover the system into the code subspace. The correction procedure consists solely of local operations and does not introduce any additional entanglement.

In the logical measurement stage (Fig.~\ref{fig3} (c)), both parties implement a set of incompatible logical measurement operators in Eq.~\eqref{measurement operator} on their logical qubits. The measurement circuit includes the phase (P) gates, Hadamard (H) gates, and rotation ($R_z(\theta)$) gates. This construction essentially assembles the phase information from all the encoded qubits to a particular physical qubit and accomplishes the single-qubit rotation on it. By choosing $\theta=0$, $\pi/2$, $\pi/4$ and $-\pi/4$, the parties can realize the measurement bases $A_0$, $A_1$, $B_0$, and $B_1$, respectively. It is noticed that within the code subspace $\mathcal{H}_{\mathrm{code}}$, the functions of $\bar{Z}$ and $\bar{X}$ coincide with the operators $Z_L=Z^{\otimes n}$ and $X_L=X^{\otimes n}$ acting on the physical Hilbert space $\mathcal{H}_{\mathrm{phy}}$, respectively, for $\bar{Z}$ and $Z_L$ ($\bar{X}$ and $X_L$) share the same eigenstates $\{|0_L\rangle, |1_L\rangle\}$ ($\{|+_L\rangle, |-_L\rangle\}$) within the code subspace. Therefore, after the error correction stage, the logical observable $\bar{Z}$ ($\bar{X}$) can be realize by the product of the physical observables $Z$ ($X$) on the physical qubits \cite{QEC1}. The detailed formula derivations of the whole circuit are seen in the Supplementary Material.

CNOT gate is critical for implementing the logical Bell test circuit. We assume that each CNOT gate has the successful probability of $p_{\mathrm{Cs}}$. Adopting a conservative model where any CNOT failure leads to a logical failure, the observed practical logical CHSH value becomes
\begin{equation}\label{SP}
S_p = p_{\mathrm{Cs}}^{2(N_H+N_M)} S_{\mathrm{CHSH}},
\end{equation}
where $N_H=\|H\|_0$ is the number of CNOT gates required for one round of stabilizer syndrome extraction, and $N_M=2(n-1)$ is the number of CNOT gates required by the logical measurement circuit. In the logical Bell test based on the $[[3,1,1]]$ repetition code, we can obtain $N_{\mathrm{tot}}=2(N_H+N_M)=16$ and a threshold of $p_{\mathrm{Cs}}=97.86\%$ is required for retaining the logical Bell nonlocality. In experiments, the CNOT gate can be equivalently constructed from the controlled-phase (CZ) gates by applying single-qubit H rotations on the target qubit. The experimental realization of CNOT gates with success probability above 99.5\% have been achieved in neutral-atom systems \cite{CNOT1,CNOT2}. Single-qubit gates together with the CNOT gate constitute a universal quantum gate set, enabling the implementation of arbitrary quantum gate operations, including $H=e^{i\frac{\pi}{2}} R_y (\frac{\pi}{2})R_z (\pi)$ gate and $P=e^{i\frac{\pi}{4}} R_z (\frac{\pi}{2})$. $R_y (\theta)$ rotations can be implemented by tuning the phase of the microwave field, while $R_z (\theta)$ operations can be realized through compositions of the rotations about the $x$ and $y$ axes \cite{Gate1}. High-fidelity single-qubit gate operations have been experimentally demonstrated in the neutral-atom systems \cite{Gate2,Gate5,Gate4} and molecular magnet systems \cite{Gate3}. In this way, the logical Bell test is achievable under current experimental conditions.

In summary, quantum nonlocality constitutes a fundamental feature of quantum physics and drives the development of the DI quantum information technologies, such as DI quantum communication, DI random number generation and self-testing. Existing physical quantum nonlocality is inevitably affected by environmental decoherence, which limit the performance of its future DI applications. In this work, we construct a general logical Bell inequality which is universally defined for the logical entanglement of various QEC codes. We prove that the violation of the logical Bell inequality indicates the global nonlocal feature of the logical system in the stabilizer coding subspace. The inequality violation depends sensitively on the noise model and the structure of the QEC code, and display superior noise robustness. For example, comparing with the physical Bell inequality violation, the fidelity thresholds for the logical Bell inequality violation based on the $[[3,1,1]]$ and $[[7,1,1]]$ repetition codes under the bit-flip error model can be reduced from 82.8\% to 73.10\% and 66.35\%, respectively, which can increase DI QKD's bit-flip noise threshold from 10.64\% to 14.42\% and 23.36\%, respectively. The stabilizer-based framework considered here is not restricted to the bipartite case and can be used in the multipartite logical entanglement case in principle.  Finally, a feasible logical Bell test implementation circuit based on the $[[3,1,1]]$ repetition code is presented. This work provides a feasible avenue for unlocking robust logical Bell nonlocality in scalable logical quantum systems, and facilitates its applications in future scalable logical quantum network.

\textbf{Acknowledgement} This work was supported by the National Natural Science Foundation of China under Grants Nos. 92365110 and 12574393.

\end{document}


\title{Supplementary Material\\Robust logical Bell nonlocality based on quantum error correction codes}
	
\author{Qi Zhang$^{1,2}$, Jia-Wei Ying$^{3}$, Cheng Liu$^{1}$,  Lan Zhou$^{1}$\footnote{Email address: zhoul727@hznu.edu.cn}, Yu-Bo Sheng$^{1,3}$\footnote{Email address: shengyb@njupt.edu.cn}}
\address{$^1$School of Physics, Hangzhou Normal University, Hangzhou, Zhejiang 311121, China\\
$^2$College of Science, Nanjing University of Posts and Telecommunications, Nanjing, Jiangsu 210023, China\\
$^3$College of Electronic and Optical Engineering and College of Flexible Electronics (Future Technology), Nanjing University of Posts and Telecommunications, Nanjing, Jiangsu 210023, China\\
}
\date{\today}

\maketitle

\section{Noise robustness of physical Bell nonlocality and logical Bell nonlocality based on various quantum error correction codes}
\subsection{The physical CHSH inequality and its noise robustness}

\begin{figure}
\includegraphics[width=0.45\textwidth]{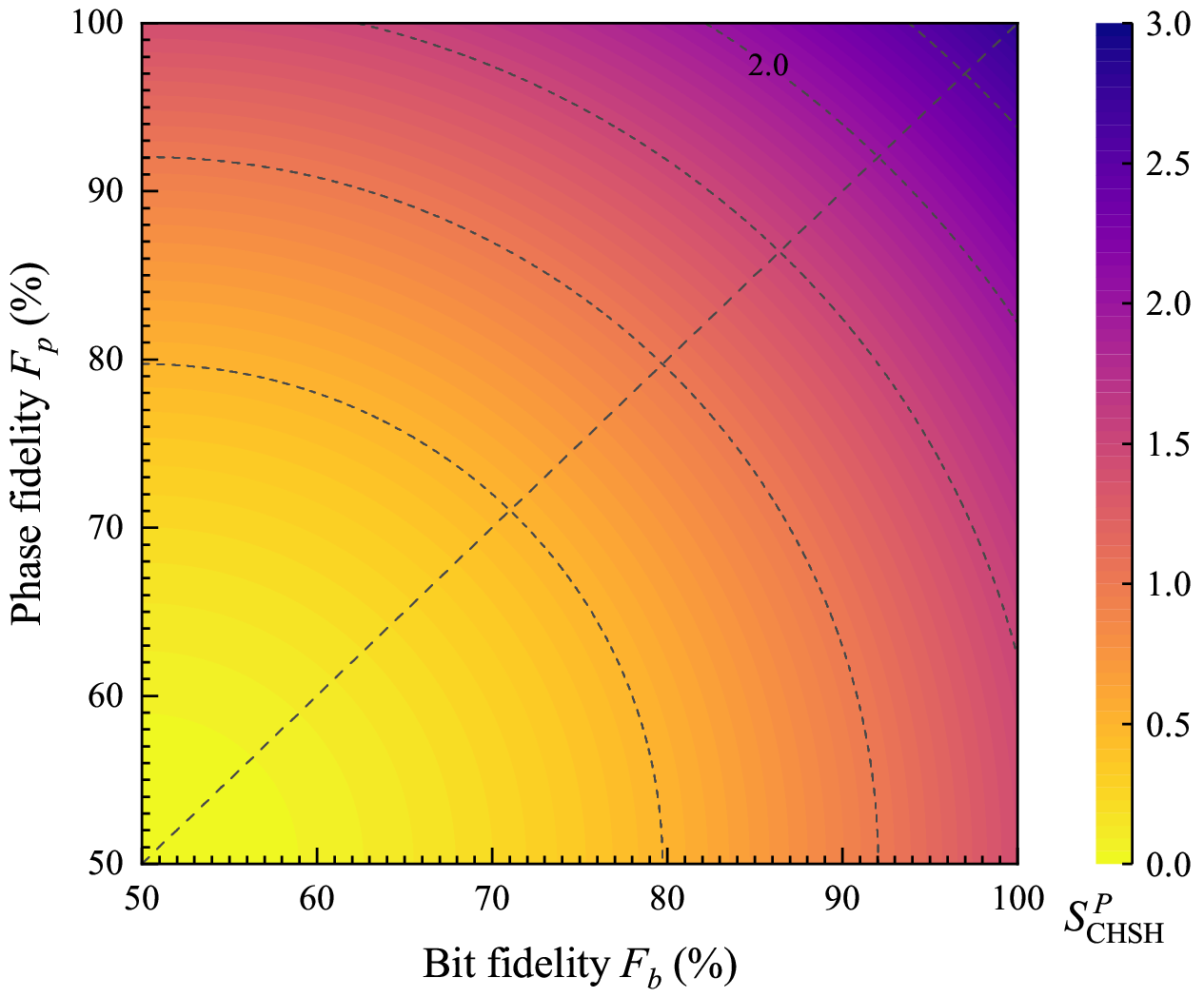}
\caption{The physical CHSH polynomial $S_{CHSH}^{P}$ altered with the physical bit fidelity $F_b$ and the physical phase fidelity $F_p$.}
\label{figS1}
\end{figure}

For the target physical Bell state $|\Phi^+\rangle= \frac{1}{\sqrt{2}}\left(|00\rangle+|11\rangle\right)$, the physical CHSH operator is defined as
\begin{eqnarray}\label{CHSH operator}
\mathcal{B_P}=A_0\otimes B_0 + A_0\otimes B_1 + A_1\otimes B_0 - A_1\otimes B_1,
\end{eqnarray}
where $A_0=Z=|0\rangle\langle0|-|1\rangle\langle1|$, $A_1=X=|0\rangle\langle1|+|1\rangle\langle0|$, $B_0=\frac{Z+X}{\sqrt{2}}$, and $B_1=\frac{Z-X}{\sqrt{2}}$.

Considering the general noise model, we assume that each physical qubit independently undergoes bit error and phase error, characterized by fidelities $F_b$ and $F_p$, respectively. Under this noise model, the target physical Bell state $|\Phi^+\rangle$ degrades into a mixed state as
\begin{eqnarray}\label{physical qubit mixed state}
\rho = F_b^B F_p^B |\Phi^+\rangle\langle\Phi^+| + F_b^B (1 - F_p^B) |\Phi^-\rangle\langle\Phi^-|+ (1 - F_b^B) F_p^B |\Psi^+\rangle\langle\Psi^+| + (1 - F_b^B)(1 - F_p^B) |\Psi^-\rangle\langle\Psi^-|,
\end{eqnarray}
where $|\Phi^{-}\rangle = \frac{1}{\sqrt{2}}\left(|00\rangle-|11\rangle\right)$ and $|\Psi^{\pm}\rangle = \frac{1}{\sqrt{2}}\left(|01\rangle+|10\rangle\right)$. Here, $F_b^B$ and $F_p^B$ denote the effective bit fidelity and phase fidelity of the Bell state with the forms of
\begin{eqnarray}\label{single-qubit fidelity}
F_b^B = F_b^2 + (1-F_b)^2, \qquad F_p^B = F_p^2 + (1-F_p)^2.
\end{eqnarray}
Therefore, the physical CHSH polynomial $S_{CHSH}^{P} $ is
\begin{eqnarray}\label{CHSH polynomial}
S_{CHSH}^{P} = \mathrm{Tr}(\mathcal{B_P} \rho) = 2\sqrt{2}\left( F_b^B F_p^B - (1-F_b^B)(1-F_p^B) \right) = 2\sqrt{2}(F_b^B + F_p^B - 1).
\end{eqnarray}

In Fig.~\ref{figS1}, we demonstrate the value of $S_{CHSH}^{P} $ altered with $F_b$ and $F_p$ under the general noise model. A violation of the CHSH inequality ($S_{CHSH}^{P}>2$) is obtained only when both $F_b$ and $F_p$ are higher than 82.20\%.

\subsection{The noise robustness of the logical Bell nonlocality based on the $[[3,1,1]]$ phase code}
\begin{figure}
\includegraphics[width=0.45\textwidth]{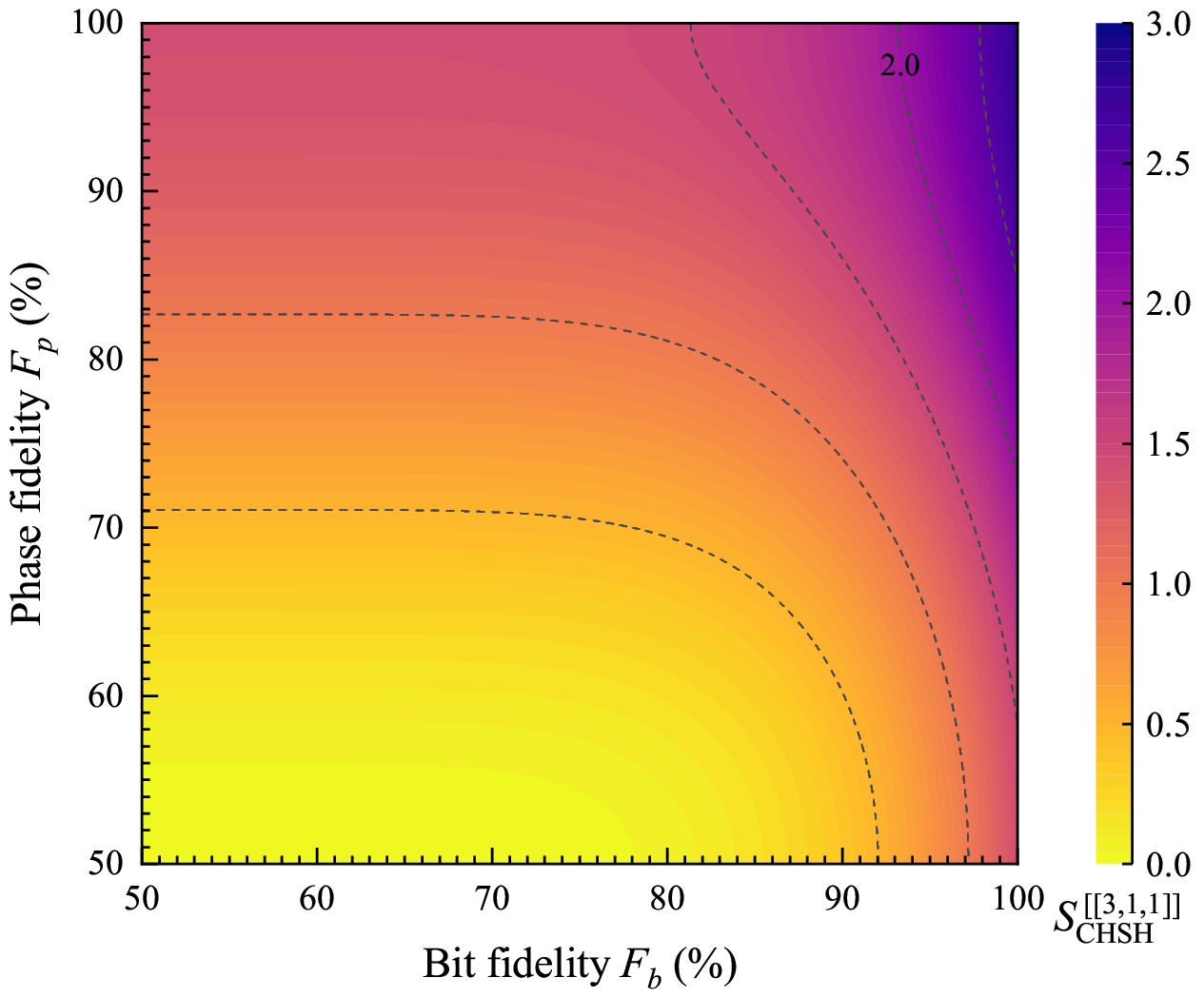}
\caption{The logical CHSH polynomial $S_{CHSH}^{[[3,1,1]]}$ altered with the physical bit fidelity $F_b$ and the physical phase fidelity $F_p$.}
\label{figS2}
\end{figure}

The $[[3,1,1]]$ phase code encodes three physical qubits into one logical qubit. The stabilizer group $\mathcal{S}^{[[3,1,1]]}=\{S_{1}^{[[3,1,1]]}, S_{2}^{[[3,1,1]]}\}$ is generated by
\begin{eqnarray}\label{phase code stabilizer}
S_{1}^{[[3,1,1]]} = X_1 \otimes X_2 \otimes I_3,\qquad S_{2}^{[[3,1,1]]} = I_1 \otimes X_2 \otimes X_3.
\end{eqnarray}
The corresponding logical basis states can be written as $|0_{L}^{[[3,1,1]]}\rangle = |+++\rangle$ and $|1_{L}^{[[3,1,1]]}\rangle = |---\rangle$, where $|+\rangle = \frac{|0\rangle + |1\rangle}{\sqrt{2}}$ and $|-\rangle = \frac{|0\rangle - |1\rangle}{\sqrt{2}}$.

The $[[3,1,1]]$ phase code can only correct the single-qubit phase error, but cannot correct the bit error. After the error correction, the phase fidelity and bit fidelity of a logical qubit are given by
\begin{eqnarray}\label{phase code logical qubit}
F_{pL}^{[[3,1,1]]}=F_p^3+3F_p^2(1-F_p), \qquad F_{bL}^{[[3,1,1]]}=F_b^3+3F_b(1-F_b)^2,
\end{eqnarray}
where $F_p$ and $F_b$ are the physical phase fidelity and bit fidelity, respectively.

In this way, under the general noise model, the initial logical Bell state $|\Phi_L^{+}\rangle$ is degraded into a mixed state as
\begin{eqnarray}\label{Phase code mixed state}
\rho_L^{[[3,1,1]]} &=& F_{b_{[[3,1,1]]}}^{B} F_{p_{[[3,1,1]]}}^{B} |\Phi_L^+ \rangle_{[[3,1,1]]}\langle \Phi_L^+| + F_{b_{[[3,1,1]]}}^{B}(1-F_{p_{[[3,1,1]]}}^{B}) |\Phi_L^- \rangle_{[[3,1,1]]}\langle \Phi_L^-| \nonumber\\
&+& (1-F_{b_{[[3,1,1]]}}^{B})F_{p_{[[3,1,1]]}}^{B} |\Psi_L^+ \rangle_{[[3,1,1]]}\langle \Psi_L^+| + (1-F_{b_{[[3,1,1]]}}^{B})(1-F_{p_{[[3,1,1]]}}^{B}) |\Psi_L^- \rangle_{[[3,1,1]]}\langle \Psi_L^-|,
\end{eqnarray}
where $|\Phi_L^{\pm} \rangle_{[[3,1,1]]}=\frac{1}{\sqrt{2}}\left(|0_{L}^{[[3,1,1]]}\rangle|0_{L}^{[[3,1,1]]}\rangle\pm |1_{L}^{[[3,1,1]]}\rangle|1_{L}^{[[3,1,1]]}\rangle\right)$ and $|\Psi_L^{\pm} \rangle_{[[3,1,1]]}=\frac{1}{\sqrt{2}}\left(|0_{L}^{[[3,1,1]]}\rangle|1_{L}^{[[3,1,1]]}\rangle\pm |1_{L}^{[[3,1,1]]}\rangle|0_{L}^{[[3,1,1]]}\rangle\right)$. Here, $F_{b_{[[3,1,1]]}}^{B}$ and $F_{p_{[[3,1,1]]}}^{B}$ denote the effective bit fidelity and phase fidelity of the target logical Bell state after the error correction as
\begin{eqnarray}\label{Phase code Bell state}
F_{b_{[[3,1,1]]}}^{B}={F_{bL}^{[[3,1,1]]}}^2+\left(1-F_{bL}^{[[3,1,1]]}\right)^2,\qquad F_{p_{[[3,1,1]]}}^{B}={F_{pL}^{[[3,1,1]]}}^2+\left(1-F_{pL}^{[[3,1,1]]}\right)^2.
\end{eqnarray}

Therefore, the logical CHSH polynomial $S_{CHSH}^{[[3,1,1]]}$ is
\begin{eqnarray}\label{311 logic CHSH polynomial}
S_{CHSH}^{[[3,1,1]]}= \mathrm{Tr}(\mathcal{B}_L \rho_L^{[[3,1,1]]}) &=& 2\sqrt{2}\left(F_{b_{[[3,1,1]]}}^{B}F_{p_{[[3,1,1]]}}^{B}-(1-F_{b_{[[3,1,1]]}}^{B})(1-F_{p_{[[3,1,1]]}}^{B})\right) \nonumber\\
&=& 2\sqrt{2}\left(F_{b_{[[3,1,1]]}}^{B}+F_{p_{[[3,1,1]]}}^{B}-1\right).
\end{eqnarray}

As shown in Fig.~\ref{figS2}, a violation of the logical CHSH inequality ($S_{CHSH}^{[[3,1,1]]}>2$) is observed only when the phase fidelity satisfies $F_p>73.10\%$ and $F_b>93.20\%$. Comparing with the physical Bell nonlocality, the logical Bell nonlocality based on the $[[3,1,1]]$ phase code is more robust against the phase error but less robust against the bit error. This behavior directly reflects the intrinsic asymmetry in the error correction capability of the $[[3,1,1]]$ phase code.

\subsection{The noise robustness of the logical Bell nonlocality based on $[[7,1,1]]$ repetition (phase) code}
The $[[7,1,1]]$ repetition code encodes seven physical qubits into one logical qubit. The stabilizer group $\mathcal{S}^{[[7,1,1]]}=\{S_{1}^{[[7,1,1]]}, S_{2}^{[[7,1,1]]}, S_{3}^{[[7,1,1]]}, S_{4}^{[[7,1,1]]}, S_{5}^{[[7,1,1]]}, S_{6}^{[[7,1,1]]}\}$ is generated by
\begin{eqnarray}\label{repetition code stabilizer}
S_1^{[[7,1,1]]} &=& Z_1 \otimes Z_2,\quad S_2^{[[7,1,1]]} = Z_2 \otimes Z_3,\quad S_3^{[[7,1,1]]} = Z_3 \otimes Z_4,\nonumber\\
S_4^{[[7,1,1]]} &=& Z_4 \otimes Z_5,\quad S_5^{[[7,1,1]]} = Z_5 \otimes Z_6,\quad S_6^{[[7,1,1]]} = Z_6 \otimes Z_7.
\end{eqnarray}
The corresponding logical basis states can be written as $|0_L^{[[7,1,1]]}\rangle = |0000000\rangle$ and $|1_L^{[[7,1,1]]}\rangle = |1111111\rangle$.

The $[[7,1,1]]$ repetition code can correct at most three bits of bit errors, but cannot correct the phase error. After the error correction, the bit fidelity and phase fidelity of a logical qubit are given by
\begin{eqnarray}\label{repetition code logical qubit}
F_{bL}^{[[7,1,1]]}=\sum_{n=0}^{3} \binom{7}{n} F_b^{7-n}(1-F_b)^n, \qquad F_{pL}^{[[7,1,1]]}=\sum_{n\in\{0,2,4,6\}} \binom{7}{n} F_p^{7-n}(1-F_p)^n,
\end{eqnarray}
where $F_p$ and $F_b$ are the physical phase fidelity and bit fidelity, respectively.

In this way, under the general noise model, the initial logical Bell state $|\Phi_L^{+}\rangle_{[[7,1,1]]}$ is degraded into a mixed state as
\begin{eqnarray}\label{repetition code mixed state}
\rho_L^{[[7,1,1]]} &=& F_{b_{[[7,1,1]]}}^{B} F_{p_{[[7,1,1]]}}^{B} |\Phi_L^+ \rangle_{[[7,1,1]]}\langle \Phi_L^+| + F_{b_{[[7,1,1]]}}^{B}(1-F_{p_{[[7,1,1]]}}^{B}) |\Phi_L^- \rangle_{[[7,1,1]]}\langle \Phi_L^-| \nonumber\\
&+& (1-F_{b_{[[7,1,1]]}}^{B})F_{p_{[[7,1,1]]}}^{B} |\Psi_L^+ \rangle_{[[7,1,1]]}\langle \Psi_L^+| + (1-F_{b_{[[7,1,1]]}}^{B})(1-F_{p_{[[7,1,1]]}}^{B}) |\Psi_L^- \rangle_{[[7,1,1]]}\langle \Psi_L^-|,
\end{eqnarray}
where $|\Phi_L^{\pm} \rangle_{[[7,1,1]]}=\frac{1}{\sqrt{2}}\left(|0_{L}^{[[7,1,1]]}\rangle|0_{L}^{[[7,1,1]]}\rangle\pm |1_{L}^{[[7,1,1]]}\rangle|1_{L}^{[[7,1,1]]}\rangle\right)$ and $|\Psi_L^{\pm} \rangle_{[[7,1,1]]}=\frac{1}{\sqrt{2}}\left(|0_{L}^{[[7,1,1]]}\rangle|1_{L}^{[[7,1,1]]}\rangle\pm |1_{L}^{[[7,1,1]]}\rangle|0_{L}^{[[7,1,1]]}\rangle\right)$. Here, $F_{b_{[[7,1,1]]}}^{B}$ and $F_{p_{[[7,1,1]]}}^{B}$ denote the effective bit fidelity and phase fidelity of the target logical Bell state after the error correction as
\begin{eqnarray}\label{repetition code Bell state}
F_{b_{[[7,1,1]]}}^{B}={F_{bL}^{[[7,1,1]]}}^2+\left(1-F_{bL}^{[[7,1,1]]}\right)^2,\qquad F_{p_{[[7,1,1]]}}^{B}={F_{pL}^{[[7,1,1]]}}^2+\left(1-F_{pL}^{[[7,1,1]]}\right)^2.
\end{eqnarray}
Therefore, the logical CHSH polynomial $S_{CHSH}^{[[7,1,1]]}$ is
\begin{eqnarray}\label{711 logic CHSH polynomial}
S_{CHSH}^{[[7,1,1]]}= \mathrm{Tr}(\mathcal{B}_L \rho_L^{[[7,1,1]]}) &=& 2\sqrt{2}\left(F_{b_{[[7,1,1]]}}^{B}F_{p_{[[7,1,1]]}}^{B}-(1-F_{b_{[[7,1,1]]}}^{B})(1-F_{p_{[[7,1,1]]}}^{B})\right) \nonumber\\
&=& 2\sqrt{2}\left(F_{b_{[[7,1,1]]}}^{B}+F_{p_{[[7,1,1]]}}^{B}-1\right).
\end{eqnarray}

\begin{figure}
\includegraphics[width=0.45\textwidth]{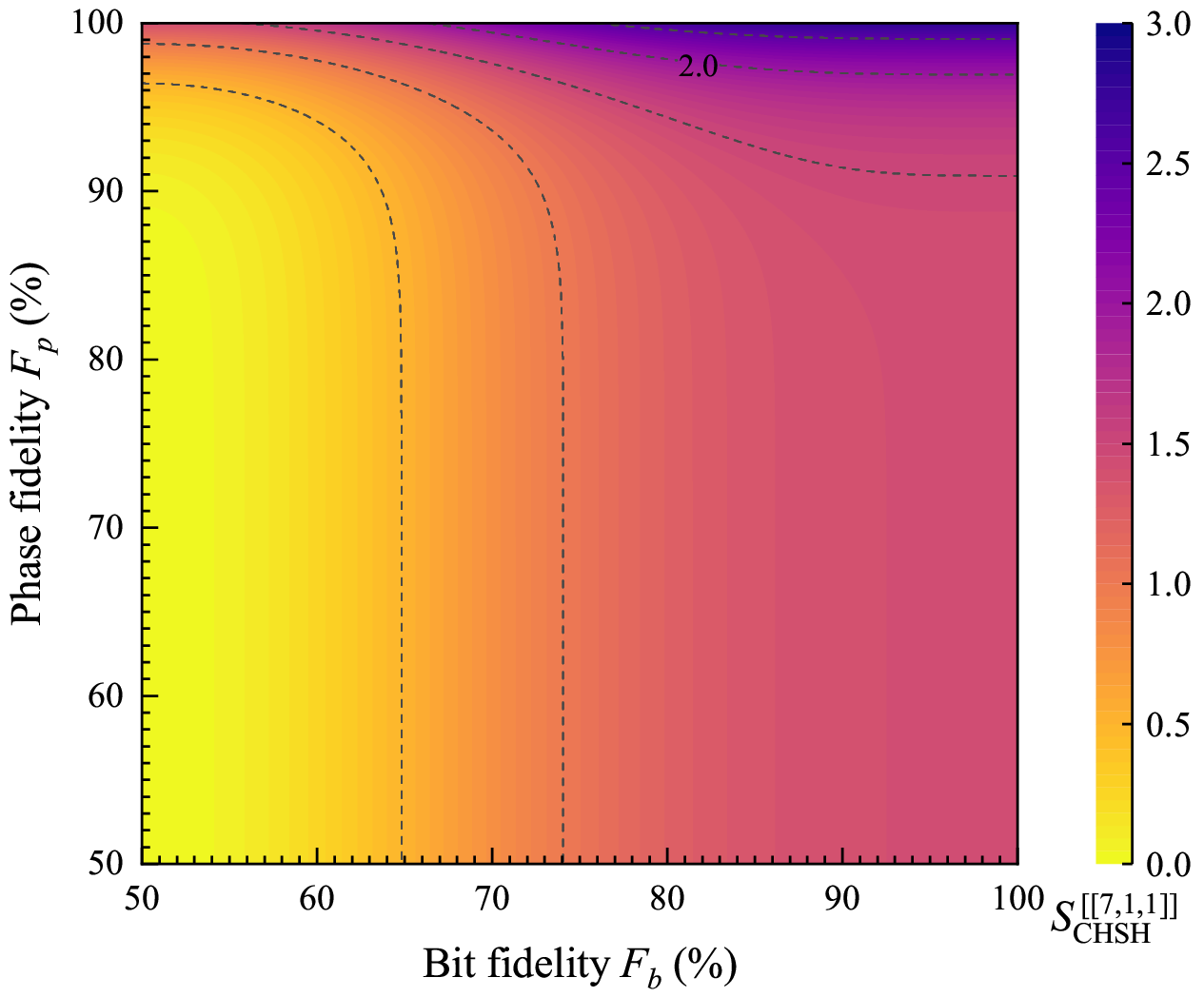}
\caption{The logical CHSH polynomial $S_{CHSH}^{[[7,1,1]]}$ altered with the physical bit fidelity $F_b$ and the physical phase fidelity $F_p$.}
\label{figS3}
\end{figure}

As shown in Fig.~\ref{figS3}, $S_{CHSH}^{[[7,1,1]]}>2$ is observed only when $F_b>66.35\%$ and $F_p>96.95\%$. Compared with the $[[3,1,1]]$ repetition code, the logical Bell nonlocality based on the $[[7,1,1]]$ repetition code exhibits substantially enhanced robustness against bit errors, while becoming more sensitive to phase errors. Correspondingly, the $[[7,1,1]]$ phase code displays the opposite noise response ($F_p>66.35\%$ and $F_b>96.95\%$), which we do not discuss further here.

\subsection{The noise robustness of the logical Bell nonlocality based on $[[7,1,3]]$ CSS code}
The $[[7,1,3]]$ CSS code encodes seven physical qubits into one logical qubit. The stabilizer group $\mathcal{S}^{[[7,1,3]]}=\{S_{1}^{[[7,1,3]]}, S_{2}^{[[7,1,3]]}, S_{3}^{[[7,1,3]]}, S_{4}^{[[7,1,3]]}, S_{5}^{[[7,1,3]]}, S_{6}^{[[7,1,3]]}\}$ is generated by
\begin{eqnarray}\label{repetition code stabilizer}
S_1^{[[7,1,3]]} &=& I_1 \otimes X_2 \otimes X_3 \otimes X_4 \otimes X_5 \otimes I_6 \otimes I_7,\nonumber\\
S_2^{[[7,1,3]]} &=& I_1 \otimes Z_2 \otimes Z_3 \otimes Z_4 \otimes Z_5 \otimes I_6 \otimes I_7,\nonumber\\
S_3^{[[7,1,3]]} &=& X_1 \otimes I_2 \otimes X_3 \otimes X_4 \otimes I_5 \otimes X_6 \otimes I_7,\nonumber\\
S_4^{[[7,1,3]]} &=& Z_1 \otimes I_2 \otimes Z_3 \otimes Z_4 \otimes I_5 \otimes Z_6 \otimes I_7,\nonumber\\
S_5^{[[7,1,3]]} &=& X_1 \otimes X_2 \otimes I_3 \otimes X_4 \otimes I_5 \otimes I_6 \otimes X_7,\nonumber\\
S_6^{[[7,1,3]]} &=& Z_1 \otimes Z_2 \otimes I_3 \otimes Z_4 \otimes I_5 \otimes I_6 \otimes Z_7.
\end{eqnarray}
The corresponding logical basis states can be written as
\begin{eqnarray}\label{CSS code state}
|0_L^{[[7,1,3]]}\rangle&=&\frac{1}{2\sqrt{2}}\left(|0000000\rangle+|1010101\rangle+|0110011\rangle+|1100110\rangle +|0001111\rangle+|1011010\rangle+|0111100\rangle+|1101001\rangle\right), \nonumber\\
|1_L^{[[7,1,3]]}\rangle&=&\frac{1}{2\sqrt{2}}\left(|1111111\rangle+|0101010\rangle+|1001100\rangle+|0011001\rangle +|1110000\rangle+|0100101\rangle+|1000011\rangle+|0010110\rangle\right).\nonumber\\
\end{eqnarray}

The $[[7,1,3]]$ CSS code can correct any single physical qubit error. Even when the errors exceed its strict error correction capability, the CSS code still retains a certain error tolerance. For instance, when any three physical qubits simultaneously suffer bit (phase) errors, 28 patterns out of the 35 possible error patterns can still be successfully corrected and projected back into the correct logical coding subspace. After the error correction, the bit fidelity and phase fidelity of a logical qubit are given by
\begin{eqnarray}\label{CSS code logical qubit}
F_{bL}^{[[7,1,3]]} &=& \binom{7}{0} F_b^{7} + \binom{7}{1} F_b^{6}(1-F_b) + \frac{28}{35} \binom{7}{3} F_b^{4}(1-F_b)^3 + \frac{7}{35} \binom{7}{4} F_b^{3}(1-F_b)^4 + \binom{7}{5} F_b^{2}(1-F_b)^5 \nonumber\\
&=& F_b^{7} + 7 F_b^{6}(1-F_b) + 28 F_b^{4}(1-F_b)^3 + 7 F_b^{3}(1-F_b)^4 + 21F_b^{2}(1-F_b)^5, \nonumber\\
F_{pL}^{[[7,1,3]]} &=& \binom{7}{0} F_p^{7} + \binom{7}{1} F_p^{6}(1-F_p) + \frac{28}{35} \binom{7}{3} F_p^{4}(1-F_p)^3 + \frac{7}{35} \binom{7}{4} F_p^{3}(1-F_p)^4 + \binom{7}{5} F_p^{2}(1-F_p)^5 \nonumber\\
&=& F_p^{7} + 7 F_p^{6}(1-F_p) + 28 F_p^{4}(1-F_p)^3 + 7 F_p^{3}(1-F_p)^4 + 21F_p^{2}(1-F_p)^5,
\end{eqnarray}
where $F_p$ and $F_b$ are the physical phase fidelity and bit fidelity, respectively.

In this way, under the general noise model, the initial logical Bell state $|\Phi_L^{+}\rangle_{[[7,1,3]]}$ is degraded into a mixed state as
\begin{eqnarray}\label{713 code mixed state}
\rho_L^{[[7,1,3]]} &=& F_{b_{[[7,1,3]]}}^{B} F_{p_{[[7,1,3]]}}^{B} |\Phi_L^+ \rangle_{[[7,1,3]]}\langle \Phi_L^+| + F_{b_{[[7,1,3]]}}^{B}(1-F_{p_{[[7,1,3]]}}^{B}) |\Phi_L^- \rangle_{[[7,1,3]]}\langle \Phi_L^-| \nonumber\\
&+& (1-F_{b_{[[7,1,3]]}}^{B})F_{p_{[[7,1,3]]}}^{B} |\Psi_L^+ \rangle_{[[7,1,3]]}\langle \Psi_L^+| + (1-F_{b_{[[7,1,3]]}}^{B})(1-F_{p_{[[7,1,3]]}}^{B}) |\Psi_L^- \rangle_{[[7,1,3]]}\langle \Psi_L^-|,
\end{eqnarray}
where $|\Phi_L^{\pm} \rangle_{[[7,1,3]]}=\frac{1}{\sqrt{2}}\left(|0_{L}^{[[7,1,3]]}\rangle|0_{L}^{[[7,1,3]]}\rangle\pm |1_{L}^{[[7,1,3]]}\rangle|1_{L}^{[[7,1,3]]}\rangle\right)$ and $|\Psi_L^{\pm} \rangle_{[[7,1,3]]}=\frac{1}{\sqrt{2}}\left(|0_{L}^{[[7,1,3]]}\rangle|1_{L}^{[[7,1,3]]}\rangle\pm |1_{L}^{[[7,1,3]]}\rangle|0_{L}^{[[7,1,3]]}\rangle\right)$. Here, $F_{b_{[[7,1,3]]}}^{B}$ and $F_{p_{[[7,1,3]]}}^{B}$ can be calculated by
\begin{eqnarray}\label{713 code Bell state}
F_{b_{[[7,1,3]]}}^{B}={F_{bL}^{[[7,1,3]]}}^2+\left(1-F_{bL}^{[[7,1,3]]}\right)^2,\qquad F_{p_{[[7,1,3]]}}^{B}={F_{pL}^{[[7,1,3]]}}^2+\left(1-F_{pL}^{[[7,1,3]]}\right)^2.
\end{eqnarray}

Therefore, the logical CHSH polynomial $S_{CHSH}^{[[7,1,3]]}$ is
\begin{eqnarray}\label{713 logic CHSH polynomial}
S_{CHSH}^{[[7,1,3]]}= \mathrm{Tr}(\mathcal{B}_L \rho_L^{[[7,1,3]]}) &=& 2\sqrt{2}\left(F_{b_{[[7,1,3]]}}^{B}F_{p_{[[7,1,3]]}}^{B}-(1-F_{b_{[[7,1,3]]}}^{B})(1-F_{p_{[[7,1,3]]}}^{B})\right) \nonumber\\
&=& 2\sqrt{2}\left(F_{b_{[[7,1,3]]}}^{B}+F_{p_{[[7,1,3]]}}^{B}-1\right).
\end{eqnarray}

\begin{figure}
\includegraphics[width=0.45\textwidth]{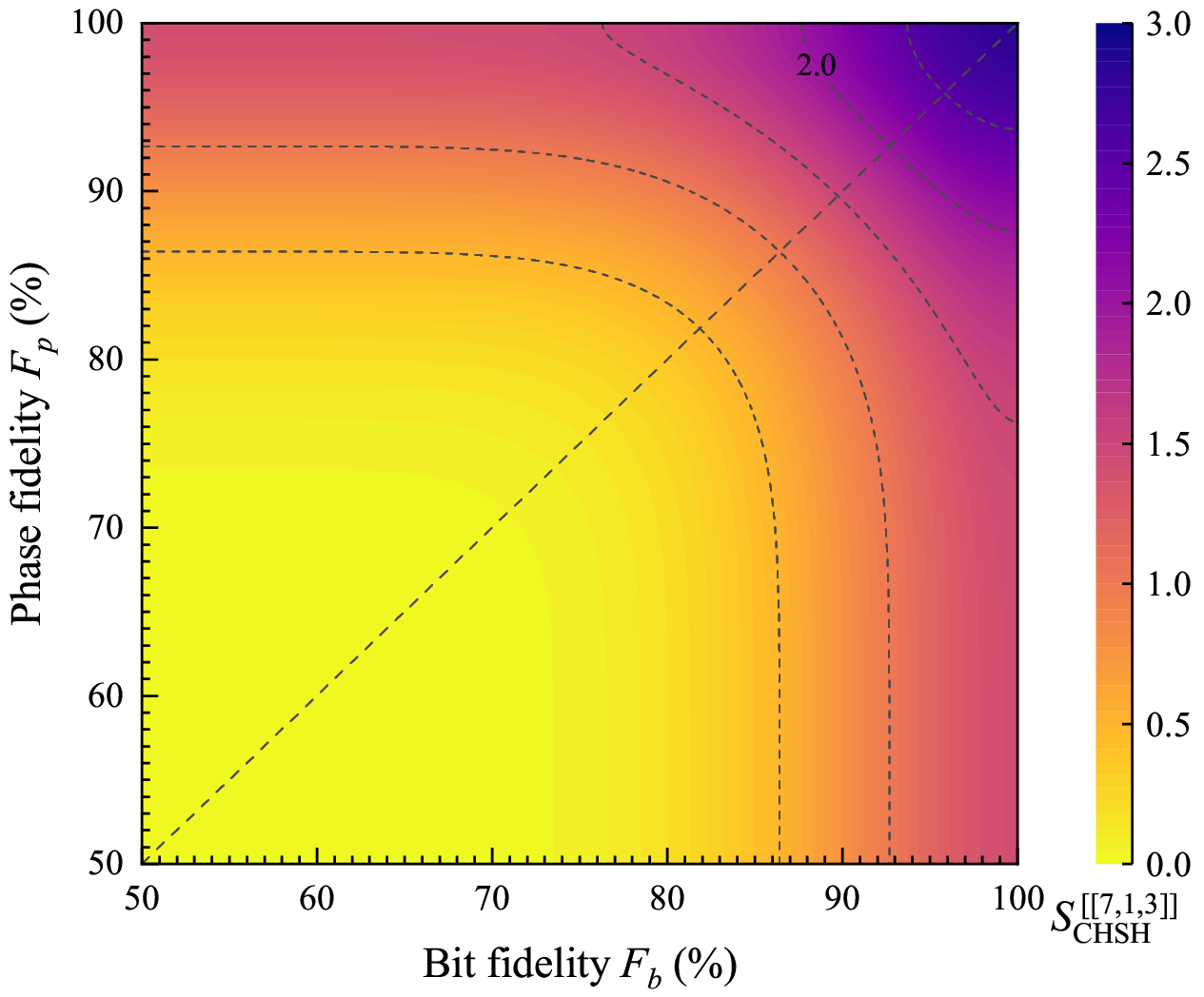}
\caption{The logical CHSH polynomial $S_{CHSH}^{[[7,1,3]]}$ altered with the physical bit fidelity $F_b$ and the physical phase fidelity $F_p$.}
\label{figS4}
\end{figure}

As shown in Fig.~\ref{figS4}, $S_{CHSH}^{[[7,1,3]]}>2$ is obtained only when both the bit fidelity and the phase fidelity are higher than 87.65\%. Compared with stabilizer codes that can only protect against a single type of error, the logical Bell nonlocality based on the $[[7,1,3]]$ CSS code exhibits symmetric and equivalent robustness against both bit error and phase error.

\subsection{The noise robustness of the logical Bell nonlocality based on $[[9,1,3]]$ Shor code}
The $[[9,1,3]]$ Shor code encodes nine physical qubits into one logical qubit. The stabilizer group $\mathcal{S}^{[[9,1,3]]}=\{S_{1}^{[[9,1,3]]}, S_{2}^{[[9,1,3]]}, S_{3}^{[[9,1,3]]}, S_{4}^{[[9,1,3]]}, S_{5}^{[[9,1,3]]}, S_{6}^{[[9,1,3]]}, S_{7}^{[[9,1,3]]}, S_{8}^{[[9,1,3]]}\}$ is generated by
\begin{eqnarray}\label{Shor code stabilizer}
S_1^{[[9,1,3]]} &=& Z_1 \otimes Z_2 \otimes I_3 \otimes I_4 \otimes I_5 \otimes I_6 \otimes I_7 \otimes I_8 \otimes I_9, \nonumber\\
S_2^{[[9,1,3]]} &=& I_1 \otimes Z_2 \otimes Z_3 \otimes I_4 \otimes I_5 \otimes I_6 \otimes I_7 \otimes I_8 \otimes I_9, \nonumber\\
S_3^{[[9,1,3]]} &=& I_1 \otimes I_2 \otimes I_3 \otimes Z_4 \otimes Z_5 \otimes I_6 \otimes I_7 \otimes I_8 \otimes I_9, \nonumber\\
S_4^{[[9,1,3]]} &=& I_1 \otimes I_2 \otimes I_3 \otimes I_4 \otimes Z_5 \otimes Z_6 \otimes I_7 \otimes I_8 \otimes I_9, \nonumber\\
S_5^{[[9,1,3]]} &=& I_1 \otimes I_2 \otimes I_3 \otimes I_4 \otimes I_5 \otimes I_6 \otimes Z_7 \otimes Z_8 \otimes I_9, \nonumber\\
S_6^{[[9,1,3]]} &=& I_1 \otimes I_2 \otimes I_3 \otimes I_4 \otimes I_5 \otimes I_6 \otimes I_7 \otimes Z_8 \otimes Z_9, \nonumber\\
S_7^{[[9,1,3]]} &=& X_1 \otimes X_2 \otimes X_3 \otimes X_4 \otimes X_5 \otimes X_6 \otimes I_7 \otimes I_8 \otimes I_9, \nonumber\\
S_8^{[[9,1,3]]} &=& I_1 \otimes I_2 \otimes I_3 \otimes X_4 \otimes X_5 \otimes X_6 \otimes X_7 \otimes X_8 \otimes X_9.
\end{eqnarray}
The corresponding logical basis states can be written as
\begin{eqnarray}\label{Shor code state}
|0_L^{[[9,1,3]]}\rangle&=&\frac{1}{2\sqrt{2}}\left(|000\rangle + |111\rangle \right)\left(|000\rangle + |111\rangle \right)\left(|000\rangle + |111\rangle \right) = |\bar{+}\rangle |\bar{+}\rangle |\bar{+}\rangle, \nonumber\\
|1_L^{[[9,1,3]]}\rangle&=&\frac{1}{2\sqrt{2}}\left(|000\rangle - |111\rangle \right)\left(|000\rangle - |111\rangle \right)\left(|000\rangle - |111\rangle \right) = |\bar{-}\rangle |\bar{-}\rangle |\bar{-}\rangle.
\end{eqnarray}
Here, the block level qubits are $|\bar{+}\rangle = \frac{|000\rangle + |111\rangle}{\sqrt{2}} = \frac{|\bar{0}\rangle + |\bar{1}\rangle}{\sqrt{2}}$ and $|\bar{-}\rangle = \frac{|000\rangle - |111\rangle}{\sqrt{2}} = \frac{|\bar{0}\rangle - |\bar{1}\rangle}{\sqrt{2}}$.

The $[[9,1,3]]$ Shor code actually embeds both the $[[3,1,1]]$ repetition code and the phase code, and can exactly correct any single physical qubit error. Similar to the $[[7,1,3]]$ CSS code, even when the noise exceeds its strict error correction capability, the code still exhibits a certain error tolerance. Therefore, after error correction, the bit fidelity and phase fidelity of the block level qubits are the same as those of the repetition code, given by
\begin{eqnarray}
B_b = F_p^3 + 3 F_p (1 - F_p)^2, \qquad B_p = F_b^3 + 3 F_b^2 (1 - F_b).
\end{eqnarray}
where $F_p$ and $F_b$ are the physical phase fidelity and bit fidelity, respectively. Therefore, the bit fidelity and phase fidelity of a single logical qubit are respectively given by
\begin{eqnarray}\label{Shor code logical qubit}
F_{bL}^{[[9,1,3]]} = B_b^3 + 3 B_b^2 (1 - B_b), \qquad F_{pL}^{[[9,1,3]]} = B_p^3 + 3 B_p (1 - B_p)^2.
\end{eqnarray}

In this way, under the general noise model, the initial logical Bell state $|\Phi_L^{+}\rangle_{[[9,1,3]]}$ is degraded into a mixed state as
\begin{eqnarray}\label{913 code mixed state}
\rho_L^{[[9,1,3]]} &=& F_{b_{[[9,1,3]]}}^{B} F_{p_{[[9,1,3]]}}^{B} |\Phi_L^+ \rangle_{[[9,1,3]]}\langle \Phi_L^+| + F_{b_{[[9,1,3]]}}^{B}(1-F_{p_{[[9,1,3]]}}^{B}) |\Phi_L^- \rangle_{[[9,1,3]]}\langle \Phi_L^-| \nonumber\\
&+& (1-F_{b_{[[9,1,3]]}}^{B})F_{p_{[[9,1,3]]}}^{B} |\Psi_L^+ \rangle_{[[9,1,3]]}\langle \Psi_L^+| + (1-F_{b_{[[9,1,3]]}}^{B})(1-F_{p_{[[9,1,3]]}}^{B}) |\Psi_L^- \rangle_{[[9,1,3]]}\langle \Psi_L^-|,
\end{eqnarray}
where $|\Phi_L^{\pm} \rangle_{[[9,1,3]]}=\frac{1}{\sqrt{2}}\left(|0_{L}^{[[9,1,3]]}\rangle|0_{L}^{[[9,1,3]]}\rangle\pm |1_{L}^{[[9,1,3]]}\rangle|1_{L}^{[[9,1,3]]}\rangle\right)$ and $|\Psi_L^{\pm} \rangle_{[[9,1,3]]}=\frac{1}{\sqrt{2}}\left(|0_{L}^{[[9,1,3]]}\rangle|1_{L}^{[[9,1,3]]}\rangle\pm |1_{L}^{[[9,1,3]]}\rangle|0_{L}^{[[9,1,3]]}\rangle\right)$. Here, the the effective bit fidelity and phase fidelity of the target logical Bell state $F_{b_{[[9,1,3]]}}^{B}$ and $F_{p_{[[9,1,3]]}}^{B}$ after the error correction can be calculated by
\begin{eqnarray}\label{913 code Bell state}
F_{b_{[[9,1,3]]}}^{B}={F_{bL}^{[[9,1,3]]}}^2+\left(1-F_{bL}^{[[9,1,3]]}\right)^2,\qquad F_{p_{[[9,1,3]]}}^{B}={F_{pL}^{[[9,1,3]]}}^2+\left(1-F_{pL}^{[[9,1,3]]}\right)^2.
\end{eqnarray}
Therefore, the logical CHSH polynomial $S_{CHSH}^{[[9,1,3]]}$ is
\begin{eqnarray}\label{913 logic CHSH polynomial}
S_{CHSH}^{[[9,1,3]]}= \mathrm{Tr}(\mathcal{B}_L \rho_L^{[[9,1,3]]}) &=& 2\sqrt{2}\left(F_{b_{[[9,1,3]]}}^{B}F_{p_{[[9,1,3]]}}^{B}-(1-F_{b_{[[9,1,3]]}}^{B})(1-F_{p_{[[9,1,3]]}}^{B})\right) \nonumber\\
&=& 2\sqrt{2}\left(F_{b_{[[9,1,3]]}}^{B}+F_{p_{[[9,1,3]]}}^{B}-1\right).
\end{eqnarray}

\begin{figure}
\includegraphics[width=0.45\textwidth]{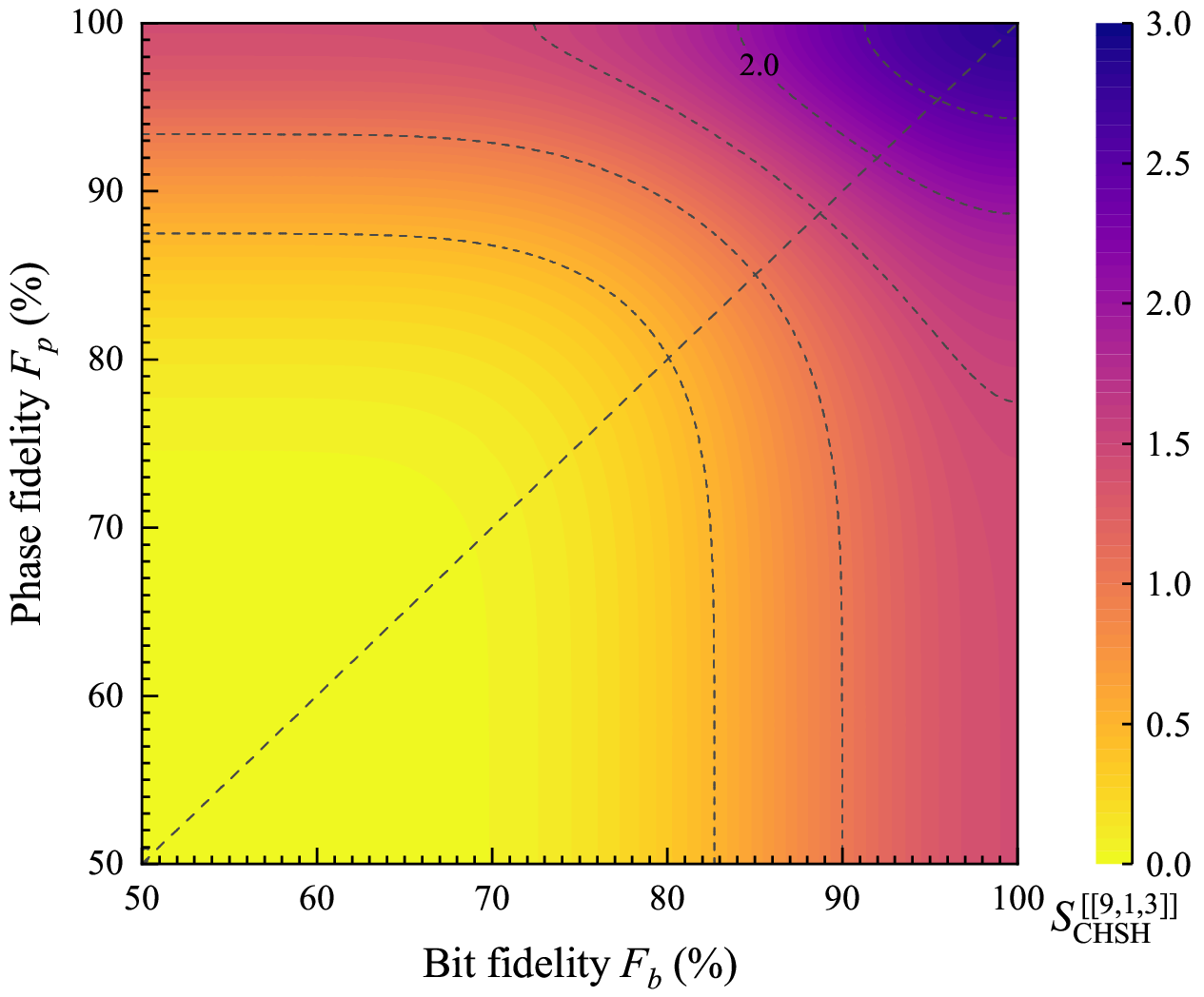}
\caption{The logical CHSH polynomial $S_{CHSH}^{[[9,1,3]]}$ altered with the physical bit fidelity $F_b$ and the physical phase fidelity $F_p$.}
\label{figS5}
\end{figure}

As shown in Fig.~\ref{figS5}, $S_{CHSH}^{[[9,1,3]]}>2$ is observed only when the bit fidelity satisfies $F_b>84.05\%$ and $F_p>88.70\%$. The logical Bell nonlocality based on the $[[9,1,3]]$ Shor code exhibits slightly higher robustness against bit error noise than that against phase error noise.

\section{Logical entanglement generation and error correction with stabilizers}
\subsection{The coding subspace of the stabilizer codes and the quantum error correction}
Stabilizer codes belong to a class of quantum error correcting codes constructed from the Pauli group. For an arbitrary $[[n,k,d]]$ stabilizer code, the $n$ qubit Pauli group is defined as \cite{QEC1}
\begin{eqnarray}
\mathcal P_n =\{\pm1,\pm i\}\otimes\{I,X,Y,Z\}^{\otimes n}.
\end{eqnarray}
Consider a set of mutually commuting Pauli operators
\begin{eqnarray}
\mathcal S = \{ S_1,S_2,\dots,S_{n-k}\},
\end{eqnarray}
satisfying $[S_i,S_j]=0$. The coding subspace of a stabilizer code is defined as the common $+1$ eigenspace of all stabilizers:
\begin{eqnarray}
\mathcal H_{\mathrm{code}} =\{|\psi\rangle: S_i|\psi\rangle=|\psi\rangle,\forall i\}.
\end{eqnarray}
Hence, every encoded logical state lies within $\mathcal H_{\mathrm{code}}$.

Suppose the encoded logical state is $|\psi_L\rangle\in\mathcal H_{\mathrm{code}}$. If an error $E\in\mathcal P_n$ occurs, the state becomes $E|\psi_L\rangle$. Since $S_i|\psi_L\rangle=|\psi_L\rangle$, the action of the stabilizer on the corrupted state is $S_iE|\psi_L\rangle$. If the error operator commutes with the stabilizer, say, $[S_i,E]=0$, we can obtain
\begin{eqnarray}
S_iE|\psi_L\rangle=ES_i|\psi_L\rangle=E|\psi_L\rangle ,
\end{eqnarray}
which corresponds to the measurement outcome $s_i=+1$. If the error operator anticommutes with the stabilizer, $\{S_i,E\}=0$, then
\begin{eqnarray}
S_iE|\psi_L\rangle=-ES_i|\psi_L\rangle=-E|\psi_L\rangle,
\end{eqnarray}
which corresponds to the measurement outcome $s_i=-1$.

Therefore, the measurement outcomes $s_i\in\{\pm1\}$ of all stabilizers collectively form the syndrome $\mathbf s=(s_1,s_2,\dots,s_{n-k})$. The syndrome characterizes how the state deviates from the coding subspace $\mathcal H_{\mathrm{code}}$. Importantly, stabilizer measurements only extract error information and do not directly measure the logical quantum state itself. Consequently, the logical information is preserved during syndrome extraction.

For each syndrome $\mathbf s$, a corresponding recovery operator $R(\mathbf s)$ is chosen such that
\begin{eqnarray}
R(\mathbf s)E|\psi_L\rangle=|\psi_L\rangle .
\end{eqnarray}
The QEC operation maps the corrupted state back into the coding subspace $\mathcal H_{\mathrm{code}}$.

\subsection{Logical entanglement generation with stabilizer}
The stabilizer not only plays a fundamental role in logical error correction procedure, but also provides an effective approach for generating logical Bell states. Here, we provide a general framework for generating logical Bell states based on the stabilizer codes \cite{Jiang}. The logical Bell state based on $[[n,k,d]]$ stabilizer code can be written as $|\Phi_L^+\rangle=\frac{1} {\sqrt{2}}\left(|0_L0_L\rangle+|1_L1_L\rangle\right)$. In principle, to generate the logical Bell state, the entanglement source first prepares $n$ physical Bell pairs $|\Phi^+\rangle=\frac{1}{\sqrt{2}}\left(|00\rangle+|11\rangle\right)$. Each particle of every physical Bell pair distributed to Alice and Bob, respectively. Then, the entanglement source performs stabilizer operations associated with the selected stabilizer code on the particles distributed to Alice and Bob, respectively. Specifically, ancillary qubits are introduced and the corresponding error syndromes are extracted through CNOT operations. The stabilizer measurements project the overall initial physical Bell state $|\Phi^+\rangle^{\otimes n}$ into a syndrome-dependent logical entangled state $E_A\otimes E_B |\Phi_L^+\rangle$, where $E_A$ and $E_B$ represent the corresponding error modes on Alice's and Bob's side, respectively. Finally, according to the extracted syndrome information, the corresponding QEC operations are applied to both sides, the shared entangled state is transformed into the target logical Bell state $|\Phi_L^+\rangle$.

In Fig.~3(a) of the main text, we show the quantum circuit for generating the logical Bell state based on the $[[3,1,1]]$ repetition code. The detailed procedure is described as follows.

First, logical entanglement source generate three physical Bell pairs
\begin{eqnarray}\label{311 entanglement generation}
|\Phi\rangle&=&|\Phi^+\rangle^{\otimes 3}=\frac{1}{\sqrt8}\left(|0_A0_B\rangle+|1_A1_B\rangle\right)\left(|0_A0_B\rangle+|1_A1_B\rangle\right)\left(|0_A0_B\rangle+|1_A1_B\rangle\right)\nonumber\\
&=&\frac{1}{\sqrt8}\big(|000\rangle_A|000\rangle_B+|001\rangle_A|001\rangle_B+|010\rangle_A|010\rangle_B+|011\rangle_A|011\rangle_B\nonumber\\
&+&|100\rangle_A|100\rangle_B+|101\rangle_A|101\rangle_B+|110\rangle_A|110\rangle_B+|111\rangle_A|111\rangle_B\big).
\end{eqnarray}

Then, the entanglement source perform stabilizer measurements on the three physical qubits on Alice's side. Assume that the measured syndrome outcome is $(s_1,s_2)=(-1,-1)$. This syndrome indicates that the first and second qubits have opposite parities, while the second and third qubits also have opposite parities. Consequently, the state in Eq.~\eqref{311 entanglement generation} is projected into the subspace spanned by $\{|010\rangle,|101\rangle\}$. Therefore, the shared entangled state collapses to
\begin{eqnarray}\label{311 error entanglement}
\frac{1}{\sqrt2} \left(|010\rangle_A|010\rangle_B+|101\rangle_A|101\rangle_B\right)=E_A\otimes E_B |\Phi_L^+\rangle,
\end{eqnarray}
where $E_A=E_B=X_2$.

For state in Eq. \eqref{311 error entanglement}, the entanglement source perform stabilizer measurements on the three physical qubits on Bob's side and yield the same syndrome outcome $(s_1,s_2)=(-1,-1)$. According to the extracted syndrome, the corresponding QEC operation $X_2$ is applied to the second qubit on Alice's and Bob's sides. After the QEC procedure, the shared state becomes $|\Phi_L^+\rangle$. Hence, the target logical Bell state is successfully generated.

It is worth noting that the stabilizer operations used here for logical Bell state generation seem to be mergeable with the subsequent logical QEC procedure. However, we emphasize that the entanglement source at this stage distributes $n$ physical Bell pairs rather than a pre-encoded logical Bell state. Directly combining these two procedures would therefore change the physical interpretation of the protocol and deviate from our original objective of generating logical Bell entanglement.

\begin{figure}
\includegraphics[width=0.2\textwidth]{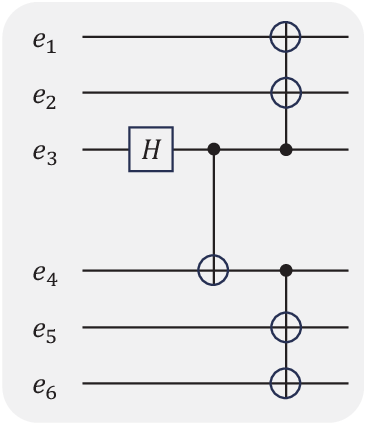}
\caption{An alternative circuit for constructing the logical Bell state $|\Phi_L^+\rangle$ based on the $[[3,1,1]]$ repetition code.}
\label{figS6}
\end{figure}

In this section, we describe a general procedure for generating logical Bell states based on stabilizer codes. This is not the only possible implementation. For example, the logical Bell state $|\Phi_L^+\rangle$ based on $[[3,1,1]]$ repetition code can also be directly constructed using Hadamard (H) and CNOT gates, as illustrated in Fig.~\ref{figS6}. The circuit employs six physical qubits $e_1-e_6$ in the state $|0\rangle$. The preparation procedure is as follows:
\begin{eqnarray}
|0\rangle_{e_1}|0\rangle_{e_2}|0\rangle_{e_3}|0\rangle_{e_4}|0\rangle_{e_5}|0\rangle_{e_6}&&\overset{H}{\longrightarrow}|0\rangle_{e_1}|0\rangle_{e_2}\frac{1}{\sqrt{2}}\left ( |0\rangle_{e_3}+|1\rangle_{e_3} \right ) |0\rangle_{e_4}|0\rangle_{e_5}|0\rangle_{e_6}\nonumber\\
&&\overset{CNOT}{\longrightarrow}\frac{1}{\sqrt{2}}\left ( |000\rangle_{e_1e_2e_3} |000\rangle_{e_4e_5e_6}+|111\rangle_{e_1e_2e_3} |111\rangle_{e_4e_5e_6}\right )=|\Phi_L^+\rangle.
\end{eqnarray}
Compared with the stabilizer projection approach introduced above, this circuit-based method directly prepares the encoded logical Bell state through gate operations, without requiring the prior distribution of physical Bell pairs and syndrome-projection procedures.

\subsection{Logical error correction with stabilizer}
For different stabilizer QEC codes, Alice and Bob need to implement different error correction circuits corresponding to the associated stabilizer measurements. As an example, Fig.~3(b) of the main text shows the error correction circuit for the $[[3,1,1]]$ repetition code.

For the $[[3,1,1]]$ repetition code, the logical basis states are defined as $|0_L\rangle=|000\rangle$ and $|1_L\rangle=|111\rangle$. An arbitrary logical state is encoded as
\begin{eqnarray}
|\psi_L\rangle=\alpha|000\rangle+\beta|111\rangle.
\end{eqnarray}
The corresponding stabilizer generators are
\begin{eqnarray}
S_1=Z_1Z_2, \qquad S_2=Z_2Z_3.
\end{eqnarray}
The coding subspace satisfies $S_1|\psi_L\rangle=|\psi_L\rangle$ and $S_2|\psi_L\rangle=|\psi_L\rangle$.

To extract the error syndrome, two ancillary qubits are introduced to measure the stabilizers $S_1$ and $S_2$, respectively. As shown in the error correction circuit of Fig.~3(b) in the main text, each ancillary qubit is initialized in the state $|0\rangle$, and CNOT gates are applied between the data qubits and the ancillary qubits according to the corresponding stabilizer operators.

After the CNOT gates, measuring the ancillary qubits in the computational basis yields the syndrome bits associated with $S_1$ and $S_2$. These syndrome outcomes indicate whether the encoded state remains inside the coding subspace $\mathcal H_{\mathrm{code}}$ or has been driven into another syndrome subspace due to a physical error.

Assume that a bit flip error occurs on the second physical qubit ($E=X_2$). The encoded state then becomes $E|\psi_L\rangle=\alpha|010\rangle+\beta|101\rangle$. After the corresponding CNOT operations, the ancillary state changes from $|00\rangle$ to $|11\rangle$. Measuring the two ancillary qubits in the $Z$ basis yields the syndrome measurement outcomes $s_1=-1$ and $s_2=-1$. This syndrome uniquely identifies a bit flip error occurring on the second physical qubit. The recovery operator is therefore chosen as $R(\mathbf s)=X_2$. Applying the recovery operation yields $X_2(\alpha|010\rangle+\beta|101\rangle)=\alpha|000\rangle+\beta|111\rangle=|\psi_L\rangle$. The system is thus restored to the coding subspace $\mathcal H_{\mathrm{code}}$, thereby completing the QEC procedure.

\section{Proof of the equivalence for logical measurement circuits}\label{Section2}
The choice of measurement basis can be mathematically understood as equivalently applying the corresponding local unitary transformation to the quantum state before measurement. Specifically, for two measurement operators $M$ and $M'$ with the same eigenvalues, there always exists a unitary operator $U$, which meets $M'=U^\dagger M U$. According to the Born rule, the measurement probability is given by the expectation value $\mathrm{Tr}(\rho M)$, and the trace is invariant under cyclic permutations, so that we can obtain
\begin{eqnarray}\label{trace invariant}
\mathrm{Tr}(\rho M') = \mathrm{Tr}(U^\dagger \rho U M).
\end{eqnarray}
This indicates that measuring the transformed state $U \rho U^\dagger$ with the fixed basis $M$ is completely equivalent to directly measuring the original state $\rho$ with the operator $M'$. Since the correlation functions depend only on the expected values of local measurement operators, switching between different measurement bases can be equivalently regarded as applying different local unitary operations to the quantum state, without affecting the Bell inequality test results.

For an arbitrary $[[n,1]]$ stabilizer code, we define $M = Z_L = Z^{\otimes n}$ and $U = e^{-i \frac{\theta}{2} Y_L}$. In this way, we can obtain
\begin{eqnarray}
M'= e^{i\frac{\theta}{2} Y_L} Z_L e^{-i \frac{\theta}{2} Y_L} = \cos\theta Z_L + \sin\theta X_L,
\end{eqnarray}
where $\theta = 0, \pi/2, \pi/4, -\pi/4$ correspond to $\bar{A}_0$, $\bar{A}_1$, $\bar{B}_0$, $\bar{B}_1$ in Eq.~(1) of the main text, respectively.

To implement the logical measurement operators $\bar{A}_0$, $\bar{A}_1$, $\bar{B}_0$, $\bar{B}_1$ in Eq.~(1) of the main text, we need to construct the corresponding logical rotation gate $U$. However, in experiments, directly applying continuously tunable rotation at the logical level is typically challenging. Therefore, efficiently switching the logical measurement basis without relying on complex logical gates, which is crucial for effectively closing the locality loophole, becomes a key issue in logical Bell tests. Fig.~3(c) in the main text illustrates an alternative approach, where the desired logical basis transformation can be realized by applying appropriate rotations on a single physical qubit. In the following part, we provide a proof that this physical level operation is fully equivalent to the target logical rotation.

Using the conjugation identity, for any unitary operator $K$ and Hermitian operator $\mathcal{C}(\theta)=e^{-i \frac{\theta}{2} Z_L}$, we have
\begin{eqnarray}
K \mathcal{C}(\theta) K^\dagger = K e^{-i \frac{\theta}{2} Z_L} K^\dagger = K \big( \cos{\frac{\theta}{2}} I - i \sin{\frac{\theta}{2}} Z_L \big) K^\dagger = e^{-i \frac{\theta}{2} (K Z_L K^\dagger)}.
\end{eqnarray}
We take $K = (S H)^{\otimes n}$, where the phase gate $S = \begin{pmatrix} 1 & 0 \\ 0 & i \end{pmatrix}$ and the Hadamard gate $H = \frac{1}{\sqrt{2}} \begin{pmatrix} 1 & 1 \\ 1 & -1 \end{pmatrix}$. Using the single qubit relation $SH Z (SH)^\dagger = Y$, we can obtain $K Z_L K^\dagger = Y_L$. Therefore, $U = e^{-i \frac{\theta}{2} Y_L} = (S H)^{\otimes n} \mathcal{C}(\theta) ((S H)^{\otimes n})^\dagger$. This shows that the logical rotation $Y_L$ can be implemented via conjugation of a rotation $Z_L$ with local phase gates and Hadamard gates on each physical qubit.

Next, we consider the construction of $\mathcal{C}(\theta) = e^{-i \frac{\theta}{2} Z_L}$. We select the $n$-th physical qubit as the target qubit and define the unitary operator as
\begin{eqnarray}
\mathcal{C}_n(\theta) = \left( \prod_{j=1}^{n-1} \text{CNOT}_{j \to n} \right) \big( I^{\otimes (n-1)} \otimes R_z(\theta) \big) \left( \prod_{j=n-1}^{1} \text{CNOT}_{j \to n} \right),
\end{eqnarray}
where $R_z(\theta) = e^{-i \frac{\theta}{2} Z}$ represents a single-qubit rotation about the $Z$ axis.

Consider an arbitrary quantum state $|x_1 x_2 \cdots x_n\rangle$, with $x_j \in {0,1}$ ($j=1,2,\cdots,n$). First, the forward chain of CNOT gates encodes the information of the first $n-1$ qubits into the last target qubit
\begin{eqnarray}
|x_1 x_2 \cdots x_n\rangle \longmapsto |x_1 \cdots (x_n \oplus x_1 \oplus \cdots \oplus x_{n-1}) \rangle.
\end{eqnarray}
In this way, a $R_z(\theta)$ rotation is applied on the target qubit, imparting a phase factor $e^{-i \frac{\theta}{2}(-1)^{x_1+\cdots+x_n}}$ to the state.

Finally, by applying the same chain of CNOT gates in the reverse order, the system can be recovered into its original quantum state with the phase factor retained. Therefore, the action of the entire operator $\mathcal{C}_n(\theta)$ on the quantum state can be expressed as
\begin{eqnarray}
\mathcal{C}_n(\theta)|x_1 x_2 \cdots x_n\rangle = e^{-i \frac{\theta}{2}(-1)^{x_1+\cdots+x_n}}|x_1 x_2 \cdots x_n\rangle.
\end{eqnarray}

Notice that the action of the logical operator $\mathcal{C}(\theta) = e^{-i \frac{\theta}{2} Z_L}$ on an arbitrary quantum state $|x_1 x_2 \cdots x_n\rangle$ can be written as
\begin{eqnarray}
\mathcal{C}(\theta)|x_1 x_2 \cdots x_n\rangle &=& (\cos\frac{\theta}{2} I - i \sin\frac{\theta}{2} Z_L)|x_1 x_2 \cdots x_n\rangle \nonumber\\
&=& e^{-i \frac{\theta}{2}(-1)^{x_1+\cdots+x_n}}|x_1 x_2 \cdots x_n\rangle.
\end{eqnarray}
Hence, we can obtain $\mathcal{C}(\theta) = \mathcal{C}_n(\theta)$. This demonstrates that any $n$-qubit logical Pauli operator can be exactly implemented using a single-qubit rotation combined with a chain of CNOT gates without the need to directly introduce multi-body interaction terms. This result provides a rigorous theoretical foundation for realizing logical rotations and switching logical measurement bases at the physical level, as employed in the main text.

\section{Effect of CNOT gate's success probability on logical Bell violations}
In the standard sequential stabilizer measurement theory, the number of CNOT gates equals to the total number of the nonzero elements in the check matrix, since the circuit structure directly mirrors the support structure of the stabilizer generators. For a stabilizer generator
\begin{eqnarray}
S_i=\bigotimes_{j=1}^{n} P_{ij}, \qquad P_{ij}\in\{I,X,Y,Z\}.
\end{eqnarray}
Its weight $\mathrm{wt}(S_i)$ is defined as the number of nontrivial Pauli operators, i.e., the number of positions where $P_{ij}\neq I$.

In the single-ancillary measurement implementation, whenever a physical qubit participates in a given stabilizer, the ancillary must be coupled to that qubit once in order to accumulate the check information onto the ancillary. Regardless of whether the operator is $X$, $Y$, or $Z$, this interaction requires exactly one CNOT gate. Therefore, each nontrivial Pauli factor corresponds to precisely one CNOT operation.

Consider arbitrary $[[n,1]]$ stabilizer code. The total number of CNOT gates required for one full round of syndrome extraction is $N_H=\sum_{i=1}^{n-1}\mathrm{wt}(S_i)$. In the binary representation with check matrix $H=(H_X|H_Z)$, each row corresponds to a stabilizer generator, and each nonzero matrix element indicates a nontrivial action of that stabilizer on a given physical qubit. Consequently,
\begin{eqnarray}
N_H=\|H\|_0,
\end{eqnarray}
where $\|H\|_0$ denotes the total number of nonzero entries in $H$, where $\|\cdot\|_0$ is $\ell_0$-norm. In other words, the number of nonzero elements in the check matrix equals to the total number of qubit couplings, and each coupling corresponds to one CNOT gate.

As discussed in Sec.~\ref{Section2}, the measurement circuit must first aggregate the parity information of all $n$ qubits onto a single target qubit. Incorporating each additional control qubit requires one CNOT operation to write its information onto the target qubit, so that a total of $n-1$ CNOT gates are necessary to complete the parity encoding. After applying a single-qubit $R_z(\theta)$ rotation on the target qubit, the encoding procedure must be reversed in order to restore the original computational basis state and preserve the amplitude structure. This uncomputation step likewise requires $n-1$ CNOT gates. Therefore, the measurement circuit additionally requires $N_M=2(n-1)$ CNOT gates.

If the entire logical operation is implemented symmetrically on two encoded blocks, the total number of CNOT gates is
\begin{eqnarray}
N_{\mathrm{tot}}=2(N_H+N_M).
\end{eqnarray}

Assuming that each CNOT gate operates independently and identically with the success probability $p_{\mathrm{Cs}}$. We adopt the conservative assumption that any single CNOT error leads to a logical failure. The success probability that all the two-qubit gates execute is
\begin{eqnarray}
P_{\mathrm{CNOT}}=p_{\mathrm{Cs}}^{2(N_H+N_M)}.
\end{eqnarray}
As a result, if the ideal logical CHSH value is $S_{\mathrm{CHSH}}$, the observed logical CHSH polynomial under this independent CNOT gate model becomes
\begin{eqnarray}
S_p=S_{\mathrm{CHSH}}P_{\mathrm{CNOT}}.
\end{eqnarray}